\documentclass[english,12pt,aps,prd,a4paper,preprintnumbers,floatfix,nofootinbib,showpacs,superscriptaddress, notitlepage]{revtex4-1} 
 \pdfoutput=1
\usepackage[usenames,dvipsnames]{color}  
\usepackage{graphicx}
\usepackage{caption}
\usepackage{subcaption}
\usepackage{amsmath}
\usepackage{amssymb}
\usepackage[colorlinks=true,citecolor=darkred,urlcolor=darkred, pdfborder={0 0 0}]{hyperref}
\usepackage[normalem]{ulem}

\usepackage[T1]{fontenc}
\usepackage[latin9]{inputenc}
\usepackage{array}
\usepackage{booktabs}
\usepackage{mathrsfs}
\usepackage{multirow}
\usepackage{tabularx}

\definecolor{darkred}{rgb}{0.6,0,0}

\definecolor{linkcolor}{rgb}{0,0,0.5}

\def\gsim{\raise0.3ex\hbox{$\;>$\kern-0.75em\raise-1.1ex\hbox{$\sim\;$}}}
\def\lsim{\raise0.3ex\hbox{$\;<$\kern-0.75em\raise-1.1ex\hbox{$\sim\;$}}}

\def\beqn#1{\begin{equation}\label{#1}}
\def\eeqn{\end{equation}}

\def\beqa#1{\begin{eqnarray}\label{#1}}
\def\eeqa{\end{eqnarray}}

\def\Z2{$\mathcal{Z_2}$}

\newcommand {\ignore}[1]{}

\newcommand{\sm}{{Standard Model }}

\def\cevns{CE$\nu$NS~}

\def\321{$\mathrm{SU(3) \otimes SU(2) \otimes U(1)}$ }

\newcommand{\AddrAHEP}{%
  AHEP Group, Institut de F\'{i}sica Corpuscular --
  CSIC/Universitat de Val\`{e}ncia, Parc Cient\'ific de Paterna.\\
 C/ Catedr\'atico Jos\'e Beltr\'an, 2 E-46980 Paterna (Valencia) - SPAIN}

\newcommand{\AddrMiranda}{%
Departamento de F\'{\i}sica, Centro de Investigaci\'on
  y de Estudios Avanzados del IPN,\\ Apartado Postal 14-740 07000 Mexico,
  Distrito Federal, Mexico}

\begin{document}

\bibliographystyle{unsrt}   

\title{\boldmath \color{BrickRed} Probing neutrino transition magnetic moments\\ with coherent elastic neutrino-nucleus scattering}

\author{O.G. Miranda}\email{omr@fis.cinvestav.mx}\affiliation{\AddrMiranda}
\author{D.K. Papoulias}\email{dipapou@ific.uv.es}\affiliation{\AddrAHEP}
\author{M. T\'ortola}\email{mariam@ific.uv.es}\affiliation{\AddrAHEP}
\author{J. W. F. Valle}\email{valle@ific.uv.es}\affiliation{\AddrAHEP}

\begin{abstract}
We explore the potential of current and next generation of coherent elastic neutrino-nucleus scattering (CE$\nu$NS) experiments in probing
neutrino electromagnetic interactions. On the basis of a thorough statistical analysis, we determine the sensitivities on each component
of the Majorana neutrino transition magnetic moment (TMM),  $\left \vert \Lambda_i \right \vert$, that follow from low-energy neutrino-nucleus 
experiments. We derive the sensitivity to neutrino TMM from the first \cevns measurement by the COHERENT
  experiment, at the Spallation Neutron Source. We also present
  results for the next phases of COHERENT using HPGe, LAr and NaI[Tl]
  detectors and for reactor neutrino experiments such as CONUS,
  CONNIE, MINER, TEXONO and RED100. The role of the CP violating
  phases in each case is also briefly discussed. We conclude that future
  \cevns experiments with low-threshold capabilities can improve
  current TMM limits obtained from  Borexino data.

\end{abstract}

\maketitle


\section{Introduction}
\label{sec:introduction}


One of the major recent milestones in particle physics has been the discovery of neutrino oscillations~\cite{Kajita:2016cak,McDonald:2016ixn,Tortola:2012te,Forero:2014bxa,deSalas:2017kay}. It implies that neutrinos are massive and, hence, new physics must exist in order to provide neutrino masses and mixings~\cite{Schechter:1980gr,Schechter:1981cv}. Massive neutrinos are expected to have non-trivial electromagnetic properties such as magnetic moments and charge radius~\cite{Schechter:1981hw,Shrock:1982sc,kayser:1982br,Nieves:1981zt,Beacom:1999wx,Maltoni:2004ei,Broggini:2012df,Giunti:2014ixa}. Here we focus on the former. Although the expected magnitude of magnetic moments is typically small, it is rather model-dependent and constitutes a precious probe of physics beyond the Standard Model (SM).

The recent observation of neutral-current coherent elastic
neutrino-nucleus scattering (CE$\nu$NS) by the COHERENT
experiment~\cite{Akimov:2017ade,Akimov:2018vzs} has given access to a
wide range of new physics opportunities. This has prompted numerous
proposals to search for physics beyond the SM~\cite{Lindner:2016wff,Billard:2018jnl,AristizabalSierra:2018eqm,Miranda:2019skf}
with a special focus on
non-standard neutrino interactions with matter~\cite{Liao:2017uzy,Dent:2017mpr,AristizabalSierra:2017joc,Denton:2018xmq,Dutta:2019eml,Coloma:2017ncl,Gonzalez-Garcia:2018dep},
sterile neutrinos~\cite{Kosmas:2017zbh,Canas:2017umu,Blanco:2019vyp},
novel
mediators~\cite{Dent:2016wcr,Farzan:2018gtr,Abdullah:2018ykz,Brdar:2018qqj}
and dark
matter~\cite{Ge:2017mcq,Ng:2017aur}. Moreover,
\cevns has been also suggested as a prominent tool towards exploring
important nuclear structure
parameters~\cite{Cadeddu:2017etk,Papoulias:2019lfi}, as well as
implications for physics
within~\cite{Cadeddu:2018izq,Ciuffoli:2018qem} and beyond the
\sm~\cite{Cadeddu:2018dux,Huang:2019ene,AristizabalSierra:2019zmy}. Very
recently, it has been emphasized the need for taking into account also
the incoherent channel of neutrino-nucleus scattering at momentum
transfers ($q$) beyond the coherency frontier, e.g. $q R_A \gg
1$~\cite{Bednyakov:2018mjd} ($R_A$ is the nuclear radius), which are
particularly relevant for neutrino floor studies at direct detection
dark matter experiments~\cite{Papoulias:2018uzy,Boehm:2018sux,Link:2019pbm}.

Here, we examine the potential of the upcoming experiments to probe
neutrino magnetic moments in their most general realization, namely
transition magnetic moments (TMMs) of Majorana
neutrinos~\cite{Schechter:1981hw}. We explore the discovery potential
of these experiments to sub-leading effects associated to neutrino
TMMs through the measurement of the \cevns event rate. 
Then, upon the work presented in~\cite{Canas:2015yoa,Tortola:2004vh,Grimus:2002vb}, we build up a  dedicated study on
low-energy neutrino-nucleus processes, in the light of current and
upcoming \cevns experiments.
In particular, we examine the potential of planned reactor neutrino experiments
   CONUS~\cite{conus},
  CONNIE~\cite{Aguilar-Arevalo:2016qen}, MINER~\cite{Agnolet:2016zir},
  TEXONO~\cite{Wong:2010zzc} and RED100~\cite{Akimov:2016xdc}, and several variants of the recent COHERENT
  experiment~\cite{Akimov:2017ade,Akimov:2018ghi,Akimov:2018vzs} at
  the Spallation Neutron Source (SNS) in probing neutrino TMMs. 
We quantify the sensitivities expected for different target materials, detector sizes, thresholds, efficiencies, exposure times
  and baseline choices. Our results are determined on the basis of a dedicated $\chi^2$ analysis that takes  into account as well 
the quenching effects, relevant for high purity sub-keV threshold detectors. We conclude that neutral-current coherent elastic 
neutrino-nucleus scattering studies at these facilities offer the capability of probing electromagnetic neutrino properties 
such as neutrino TMMs with improved sensitivities, hence providing a sensitive way to test for new physics in the neutrino sector.
Beyond the analysis of \cevns experiments, in this work we update the discussion given in Refs.~\cite{Canas:2015yoa,Tortola:2004vh,Grimus:2002vb} concerning the sensitivity of $\nu-e$ scattering to the effective neutrino magnetic moment using the solar neutrino data from the Borexino collaboration~\cite{Borexino:2017fbd}. We also briefly comment on alternatives to probe the effective neutrino magnetic moments using other neutrino sources that contribute to the neutrino floor in dark matter direct detection experiments, such as solar and geoneutrinos, as well as atmospheric and diffuse supernova neutrinos.

The paper has been organized as follows. In
  Sect.~\ref{sec:theoretical-framework}, we introduce the main
theoretical background and derive the expressions for the effective
neutrino magnetic moments corresponding to the various neutrino
sources under study. In Sect.~\ref{sec:EM-contribution-to-cevns}, we
discuss the main features associated with the relevant electromagnetic
\cevns processes, while in Sect.~\ref{sec:experimental-setup} we
define the experimental configurations and setups for the different
\cevns experiments of interest. Our results are presented in
Sect.~\ref{sec:results}. A brief discussion, including updated
constraints from the recent Borexino data, and comments on other
neutrino sources that might be relevant to the neutrino floor in dark matter
direct detection experiments are given in
Sect.~\ref{sec:other-experiments}. Finally, the main conclusions are
given in Sect.~\ref{sec:conclusions}.


\section{Theoretical framework}
\label{sec:theoretical-framework}

The effective Hamiltonian that accounts for the spin component of the
Majorana neutrino electromagnetic vertex is expressed in terms of the
electromagnetic field tensor $F_{\alpha \beta}$,
as~\cite{Schechter:1981hw,Schechter:1981cv}
\begin{equation}
H_{\text{EM}}^{\mathrm{M}} = -\frac{1}{4} \nu_L^\mathsf{T} C^{-1} \lambda \sigma^{\alpha \beta} \nu_L F_{\alpha \beta} + \mathrm{h.c.} \, ,
\label{Hamiltonian:Majorana}
\end{equation}
with $\lambda = \mu - i \epsilon $ being an antisymmetric complex
matrix ($\lambda_{\alpha \beta} = - \lambda_{\beta\alpha}$) and, hence,
$\mu^{\mathsf{T}} = -\mu$ and $\epsilon^{\mathsf{T}} = -\epsilon$ are
two imaginary matrices. Therefore, three complex or six real parameters are required to describe this object. The
corresponding Hamiltonian relevant to the Dirac neutrino case reads
\begin{equation}
H_{\text{EM}}^{\mathrm{D}} = \frac{1}{2} \bar{\nu}_R \lambda \sigma^{\alpha \beta} \nu_L F_{\alpha \beta} + \mathrm{h.c.} \, ,
\label{Hamiltonian:Dirac}
\end{equation}
where $\lambda = \mu - i \epsilon$ is a complex matrix, subject to the
hermiticity constraints $\mu = \mu^\dagger$ and $\epsilon =
\epsilon^\dagger$.
Comparing Eqs.~(\ref{Hamiltonian:Majorana}) and
(\ref{Hamiltonian:Dirac}), it becomes evident that neutrino
electromagnetic properties constitute a prime vehicle to distinguish
between the Dirac and Majorana neutrino nature. In contrast to the Dirac
case, vanishing diagonal moments are implied for  Majorana neutrinos, 
$\mu^\mathrm{M}_{ii} = \epsilon^\mathrm{M}_{ii} = 0$. 
In the simplest $\mathrm{SU(2)_L \otimes U(1)_Y}$ model, the Majorana magnetic
and electric transition moments  (with $i\ne j$) take the form~\cite{Shrock:1982sc}
\begin{equation}
\mu_{ij}^{\mathrm{M}} = -\frac{3i e G_F}{16 \pi^2 \sqrt{2}} (m_{\nu i} + m_{\nu j}) \sum_{\alpha = e, \mu, \tau}  \Im m \left[ U_{\alpha i}^{\ast} U_{\alpha j} \left( \frac{m_{l \alpha}}{M_W} \right)^2 \right] \, ,
\end{equation}
\begin{equation}
\epsilon_{ij}^{\mathrm{M}} = \frac{3i e G_F}{16 \pi^2 \sqrt{2}} (m_{\nu i} - m_{\nu j}) \sum_{\alpha = e, \mu, \tau}  \Re e \left[ U_{\alpha i}^{\ast} U_{\alpha j} \left( \frac{m_{l \alpha}}{M_W} \right)^2 \right] \, ,
\end{equation}
where $G_F$ is the Fermi coupling constant, $m_{\nu_i}$ is the mass of the neutrino mass eigenstate $\nu_i$, $U_{\alpha i}$ denote the elements of the neutrino mixing matrix, while $m_{l \alpha}$ and $M_W$ correspond to the charged lepton and W boson masses, respectively.

In this work, we will focus on the study of the Majorana  transition magnetic moment $\mu_{ij}^{\mathrm{M}}$. For simplicity, we will drop the superscript M referring to Majorana neutrinos from now on.
The effective neutrino magnetic moment, observable in a given experiment,
can be expressed in terms of the neutrino magnetic moment matrix and the 
amplitudes of positive and negative helicity states, denoted by the 
$3-$vectors $\mathfrak{a}_{+}$ and $\mathfrak{a}_{-}$, respectively. 
In the flavor basis one finds~\cite{Grimus:2000tq}
\begin{equation}
\left(\mu_\nu^{F} \right)^2 = \mathfrak{a}_{-}^\dagger \lambda^\dagger \lambda \mathfrak{a}_{-} + \mathfrak{a}_{+}^\dagger \lambda \lambda^\dagger \mathfrak{a}_{+} \, ,
\label{eq:TMM-flavor}
\end{equation}
with the magnetic moment matrix $\lambda$ ($\tilde{\lambda}$) in the flavor (mass) basis defined as 
\begin{equation}
\lambda = \left( \begin{array}{ccc}
0 & \Lambda_\tau & - \Lambda_\mu \\
- \Lambda_\tau &  0 & \Lambda_e \\
\Lambda_\mu & - \Lambda_e & 0
\end{array} \right), \qquad
\tilde{\lambda} = \left( \begin{array}{ccc}
0 & \Lambda_3 & - \Lambda_2 \\
- \Lambda_3 &  0 & \Lambda_1 \\
\Lambda_2 & - \Lambda_1 & 0
\end{array} \right) \, .
\label{NMM:matrix}
\end{equation}
In this context, the definition $\lambda_{\alpha \beta} = \varepsilon_{\alpha \beta \gamma} \Lambda_\gamma$ has been introduced, and the neutrino TMMs are represented by the complex parameters~\cite{Tortola:2004vh} 
\begin{equation}
\Lambda_{\alpha} = \vert \Lambda_\alpha \vert e^{i \zeta_\alpha}, \qquad \Lambda_{i} = \vert \Lambda_i \vert e^{i \zeta_i}\, .
\label{eq:def-lambda}
\end{equation}
The effective neutrino magnetic moment in the flavor basis, shown in Eq.~(\ref{eq:TMM-flavor}), can be
translated into the mass basis through a rotation, by using the
leptonic mixing matrix. Then, by introducing the transformations
\begin{equation}
\tilde{\mathfrak{a}}_{-} = U^\dagger \mathfrak{a}_{-}, \qquad  \tilde{\mathfrak{a}}_{+} = U^\mathsf{T} \mathfrak{a}_{+}, \qquad \tilde{\lambda} = U^\mathsf{T} \lambda U \, , 
\end{equation}
the effective neutrino magnetic moment in the mass basis takes the
form~\cite{Grimus:2002vb}
\begin{equation}
\left(\mu_\nu^{M} \right)^2 = \tilde{\mathfrak{a}}_{-}^\dagger \tilde{\lambda}^\dagger \tilde{\lambda} \tilde{\mathfrak{a}}_{-} + \tilde{\mathfrak{a}}_{+}^\dagger \tilde{\lambda} \tilde{\lambda}^\dagger \tilde{\mathfrak{a}}_{+} \, .
\label{eq:TMM-mass}
\end{equation}

\subsection{Effective neutrino magnetic moment at reactor \cevns experiments}

For \cevns studies at reactor neutrino experiments, the only non-zero parameter entering Eq.~(\ref{eq:TMM-flavor}) or
Eq.~(\ref{eq:TMM-mass}) is $\mathfrak{a}_{+}^1$, corresponding to the initial $\bar{\nu}_e$ flux. Then, in the flavor basis, the effective
Majorana TMM strength parameter relevant to reactor \cevns experiments such as CONUS, CONNIE, MINER, TEXONO and
RED100, can be cast in the form~\cite{Grimus:2002vb} 
\begin{equation}
\left(\mu_{\bar{\nu}_e, \, \text{reactor}}^F \right)^2 = |\Lambda_{\mu}|^2 + |\Lambda_{\tau}|^2 \, ,
\label{TMM-reactor-flavor}
\end{equation}
where $|\Lambda_{\mu}|$ and $|\Lambda_{\tau}|$ denote the elements of the neutrino transition magnetic moment matrix $\lambda$ describing
the corresponding conversions from the electron antineutrino to the muon and tau neutrino states, respectively. The above expression, 
in the mass basis becomes \footnote{Note that, in the symmetric parametrization of the  neutrino mixing matrix for Majorana neutrinos, where 
$U=R_{23}\left(\theta_{23} ; \phi_{23}\right) R_{13}\left(\theta_{13} ; \phi_{13}\right) R_{12}\left(\theta_{12} ; \phi_{12}\right)$ and 
$\delta_\text{CP} = \phi_{13} - \phi_{12}-\phi_{23}$~\cite{Schechter:1980gr}, all the  CP phases entering in the effective neutrino magnetic 
moment in Eq.~(\ref{eq:TMM-reactor-mass}) are of Majorana type.}~\cite{Canas:2015yoa} 
\begin{equation}
\begin{aligned}
\left(\mu_{\bar{\nu}_e, \, \text{reactor}}^M \right)^2 = & \vert \mathbf{\Lambda} \vert^2 - c_{12}^2 c_{13}^2 \vert \Lambda_1 \vert^2 - s_{12}^2 c_{13}^2 \vert \Lambda_2 \vert^2  - s_{13}^2 \vert \Lambda_3 \vert^2 \\
& - c_{13}^2 \sin 2\theta_{12}  \vert \Lambda_1 \vert \vert \Lambda_2 \vert \cos (\zeta_1 - \zeta_2) \\
& - c_{12} \sin 2\theta_{13}  \vert \Lambda_1 \vert \vert \Lambda_3 \vert \cos (\delta_{\text{CP}}+ \zeta_1 - \zeta_3) \\
& -  s_{12} \sin 2\theta_{13} \vert \Lambda_2 \vert \vert \Lambda_3 \vert \cos (\delta_{\text{CP}} + \zeta_2- \zeta_3) \, ,
\end{aligned}
\label{eq:TMM-reactor-mass}
\end{equation}
where $\vert \Lambda_i \vert$ and  $\zeta_i$ are the moduli and phases 
characterizing the neutrino TMM matrix in the mass basis, see Eq.~(\ref{eq:def-lambda}). 
We have also defined $\vert \mathbf{\Lambda} \vert^2 = \vert \Lambda_1 \vert^2 + \vert \Lambda_2 \vert^2 + \vert \Lambda_3 \vert^2$ 
and used the standard abbreviations $c_{ij} = \cos \theta_{ij}$, $s_{ij} = \sin \theta_{ij}$ for the trigonometric functions of the neutrino 
mixing angles. As usual, $\delta_{\text{CP}}$ refers to the Dirac CP phase of the leptonic mixing matrix.

The expression above can be further simplified by defining a new set of phases $\xi_i$ as the differences 
of the TMM phases: $\xi_1 = \zeta_3 - \zeta_2$, $\xi_2 = \zeta_3 -\zeta_1$ and  $\xi_3 = \zeta_1 - \zeta_2$. 
Note that  $\xi_2=\xi_1-\xi_3$ and, therefore, only two $\xi_i$ phases are independent. In the following, we will express the effective neutrino magnetic moments as a function of the $\delta_{\text{CP}}$ and $\xi_i$ phases. With this notation, the effective neutrino magnetic moment in Eq.~(\ref{eq:TMM-reactor-mass}) will be expressed as
\begin{equation}
\begin{aligned}
\left(\mu_{\bar{\nu}_e, \, \text{reactor}}^M \right)^2 = & \vert \mathbf{\Lambda} \vert^2 - c_{12}^2 c_{13}^2 \vert \Lambda_1 \vert^2  - s_{12}^2 c_{13}^2 \vert \Lambda_2 \vert^2 - s_{13}^2 \vert \Lambda_3 \vert^2 \\
& - c_{13}^2 \sin 2\theta_{12}  \vert \Lambda_1 \vert \vert \Lambda_2 \vert \cos \xi_3 \\
& - c_{12} \sin 2\theta_{13}  \vert \Lambda_1 \vert \vert \Lambda_3 \vert \cos (\delta_{\text{CP}} - \xi_2) \\
& -  s_{12} \sin 2\theta_{13} \vert \Lambda_2 \vert \vert \Lambda_3 \vert \cos (\delta_{\text{CP}} - \xi_1) \, .
\end{aligned}
\label{eq:TMM-reactor-mass2}
\end{equation}
It is interesting to notice that a degenerate case arises when the arguments of the cosine functions in Eq.~(\ref{eq:TMM-reactor-mass2}) are set to zero. Indeed, in this particular case one has~\cite{Canas:2016kfy} 
\begin{eqnarray}\label{eq:mureac2}
\left(\mu_{\bar{\nu}_e, \, \text{reactor}}^M \right)^2 &=& {|{\bf \Lambda}|^{2}} - (c_{12}c_{13}|\Lambda_{1}|+s_{12}c_{13}|\Lambda_{2}|
                   +  s_{13}|\Lambda_{3}|)^{2}\, ,
\end{eqnarray}
that will vanish for the following values of $|\Lambda_i|$ 
\begin{equation}\label{eq:mureac0}
 |\Lambda_{1}| = c_{12}c_{13}|{\bf \Lambda}|, \quad |\Lambda_{2}| = s_{12}c_{13}|{\bf \Lambda}|, \quad |\Lambda_{3}| = s_{13}|{\bf \Lambda}|. 
\end{equation}
Hence, for this special case, reactor experiments become insensitive to the neutrino magnetic moment.

\subsection{Effective neutrino magnetic moment at SNS facilities}

We now focus on DAR-$\pi$ neutrinos produced at the SNS and we express
the relevant neutrino magnetic moment accordingly. Assuming the same
proportion of delayed ($\nu_e$, $\bar{\nu}_\mu$) and prompt
($\nu_\mu$) neutrinos at the SNS, the relevant non-zero amplitudes are
$\mathfrak{a}_{-}^1 = 1$, $\mathfrak{a}_{+}^2 = 1$ and
$\mathfrak{a}_{-}^2 =1$, respectively. In Ref.~\cite{Canas:2015yoa}, the authors
explored TMMs at neutrino-electron scattering experiments and obtained
their results by assuming all relevant non-vanishing helicity
amplitudes at accelerator neutrino facilities. In contrast, in the
present work, by exploiting the fact that the SNS employs a pulsed
beam and can therefore distinguish between the prompt and delayed
neutrino fluxes~\cite{Cadeddu:2018dux,Dutta:2019eml}, we consider separately the TMMs corresponding to the
prompt and the delayed flux. For prompt neutrinos at the SNS (e.g. the
only non-vanishing entry being $\mathfrak{a}_{-}^2=1$), the effective
magnetic moment strength parameter in the flavor basis is expressed as
\begin{equation}
\left( \mu_{\nu_\mu, \, \text{prompt}}^F \right)^2 = \vert \Lambda_e \vert^2 +  \vert \Lambda_\tau \vert^2  \, ,
\label{eq:TMM-SNS-prompt-flavor}
\end{equation}
while for delayed neutrinos ($\mathfrak{a}_{-}^1=1$, $\mathfrak{a}_{+}^2=1$) we find 
\begin{equation}
\left( \mu_{\nu_e, \, \text{delayed}}^F \right)^2  = \vert \Lambda_\mu \vert^2 +  \vert \Lambda_\tau \vert^2, \quad \left( \mu_{\bar{\nu}_\mu, \, \text{delayed}}^F \right)^2  = \vert \Lambda_e \vert^2 +  \vert \Lambda_\tau \vert^2 
\label{eq:TMM-SNS-delayed-flavor}
\end{equation}
Assuming only the prompt neutrino flux at the SNS, the neutrino TMM  in the mass basis reads 
\begin{equation}
\begin{aligned}
\left( \mu_{\nu_\mu, \, \text{prompt}}^M \right)^2 &=  \left \vert \Lambda_1 \right \vert^2 \bigl[ -2 c_{12} c_{23} s_{12} s_{13} s_{23} \cos \delta_{\text{CP}}  \\ & \quad \quad \quad \quad +s_{23}^2 \left(c_{13}^2+s_{12}^2 s_{13}^2\right)+c_{12}^2 c_{23}^2 \bigr]\\
  & + \left \vert \Lambda_2 \right \vert^2 \left[ 2 c_{12} c_{23} s_{13} s_{23} s_{12} \cos \delta_{\text{CP}} +c_{23}^2 s_{12}^2+s_{23}^2 \left(c_{12}^2 s_{13}^2+c_{13}^2\right) \right] \\
  & + \left \vert \Lambda_3 \right \vert^2 \left[ c_{23}^2+s_{13}^2 s_{23}^2  \right] \\
  & + 2 \left \vert \Lambda_1 \Lambda_2 \right \vert \bigl[ c_{23} c_{12}^2 s_{13} s_{23} \cos \left(\delta_{\text{CP}} + \xi _3\right)-c_{23} s_{12}^2 s_{13} s_{23} \cos \left(\delta_{\text{CP}} - \xi _3\right) \\
 & \quad \quad \quad \quad \quad +c_{12} s_{12} \left(c_{23}^2-s_{13}^2
   s_{23}^2\right) \cos \xi _3\bigr]\\
  & + 2 \left \vert \Lambda_1 \Lambda_3 \right \vert \left[  c_{13} s_{23} \left(c_{12} s_{13} s_{23} \cos \left(\delta_{\text{CP}} -\xi _2\right)+c_{23} s_{12} \cos \xi _2\right) \right]  \\
 & + 2 \left \vert \Lambda_2 \Lambda_3 \right \vert \left[  c_{13} s_{23} \left(s_{12} s_{13} s_{23} \cos \left(\delta_{\text{CP}} -\xi _1\right)-c_{12} c_{23} \cos \xi _1\right) \right] \, .
\end{aligned}
\label{eq:TMM-SNS-prompt-mass}
\end{equation}
Similarly the effective neutrino magnetic moment relevant to delayed beam has two components, one corresponding to the $\nu_e$ beam ($\mathfrak{a}_{-}^1 = 1$)
\begin{equation}
\begin{aligned}
\left( \mu_{\nu_e, \, \text{delayed}}^M \right)^2 & = 
 \left \vert \Lambda_1  \right \vert^2  \left[c_{13}^2 s_{12}^2+s_{13}^2\right] +
  \left \vert \Lambda_2  \right \vert^2 \left[c_{12}^2 c_{13}^2+s_{13}^2\right] +
    \left \vert \Lambda_3  \right \vert^2 c_{13}^2\\
    &- \left \vert \Lambda_1 \Lambda_2 \right \vert \left[c_{13}^2  \sin (2
  \theta_{12})\cos \xi_3 \right]    -  \left \vert \Lambda_1 \Lambda_3 \right \vert     \left[ c_{12}   \sin (2\theta_{13}) \cos (\delta_\text{CP} -\xi_2)  \right]\\
  &- \left \vert \Lambda_2 \Lambda_3 \right \vert \left[ s_{12} \sin (2\theta_{13}) \cos (\delta_\text{CP} -\xi_1) \right] \, ,
\end{aligned}
\label{eq:TMM-SNS-delayed-mass-nue}
\end{equation}
and another corresponding to the  $\bar{\nu}_\mu$ beam ($\mathfrak{a}_{+}^2 = 1$)
\begin{equation}
\begin{aligned}
\left( \mu_{\bar{\nu}_\mu, \, \text{delayed}}^M \right)^2 & = \left \vert \Lambda_1 \right \vert^2 \left[ -2 c_{12} c_{23} s_{12} s_{13} s_{23} \cos \delta _{\text{CP}}+s_{23}^2 \left(c_{13}^2+s_{12}^2 s_{13}^2\right)+c_{12}^2 c_{23}^2 \right]\\
& + \left \vert \Lambda_2 \right \vert^2 \left[2 c_{12} c_{23} s_{12} s_{13} s_{23} \cos \delta _{\text{CP}}+s_{23}^2 \left(c_{13}^2 + c_{12}^2 s_{13}^2 \right)+ s_{12}^2 c_{23}^2 \right]\\
&+\left \vert \Lambda_3 \right \vert^2 \left[ \frac{1}{4} \left(2 c_{13}^2 \cos (2\theta_{23})-\cos (2\theta_{13})+3\right)\right] \\
 &+ 2 \left \vert \Lambda_1 \Lambda_2 \right \vert  \bigl[c_{23} s_{13} s_{23} \left(c_{12}^2 \cos \left(\delta _{\text{CP}}+\xi _3\right)-s_{12}^2 \cos \left(\delta _{\text{CP}}-\xi _3\right)\right) \\
  & +c_{12} c_{23}^2 s_{12} \cos \xi _3-c_{12} s_{12} s_{13}^2 s_{23}^2
   \cos \xi _3 \bigr] \\
    & +2 \left \vert \Lambda_1 \Lambda_3 \right \vert \bigl[  c_{13} s_{23} \left(c_{12} s_{13} s_{23} \cos \left(\delta _{\text{CP}}-\xi _2\right)+c_{23} s_{12} \cos \xi _2 \right) \bigr]\\
   &+ 2 \left \vert \Lambda_2 \Lambda_3 \right \vert \bigl[ c_{13} s_{23} \left(s_{12} s_{13} s_{23} \cos \left(\delta _{\text{CP}}-\xi _1\right)-c_{12} c_{23} \cos \xi _1 \right)\bigr] \, .
\end{aligned}
\label{eq:TMM-SNS-delayed-mass-numub}
\end{equation}
Notice from Eqs.(\ref{eq:TMM-reactor-mass2})  and (\ref{eq:TMM-SNS-prompt-mass})--(\ref{eq:TMM-SNS-delayed-mass-numub}) that
the factors accompanying $\left \vert \Lambda_i \right  \vert$ involve different CP phase and  mixing angle
combinations for the DAR-$\pi$ and reactor \cevns experiments. This will have a direct impact on the results presented 
in Sect.~\ref{sec:results}. 


\section{Electromagnetic contribution to \cevns}
\label{sec:EM-contribution-to-cevns}

Within the SM, the interaction of a neutrino with energy $E_\nu$ scattered coherently upon a nucleus $(A,Z)$ is theoretically
well studied~\cite{Freedman:1973yd,Papoulias:2015vxa,Bednyakov:2018mjd,Pirinen:2018gsd}. The \cevns cross section is usually expressed 
in terms of the nuclear recoil energy $T_A$, as~\cite{Kosmas:2017tsq}
\begin{equation}
\left(\frac{d \sigma}{dT_A}\right)_{\text{SM}} = \frac{G_F^2 m_A}{\pi} \left[\mathcal{Q}_V^2 \left(1 - \frac{m_A T_A}{2 E_\nu^2}\right) + \mathcal{Q}_A^2 \left(1 + \frac{m_A T_A}{2 E_\nu^2}\right) \right] F^2(Q^2) \, ,
\label{eq:xsec-cevns}
\end{equation}
where $m_A$ denotes the nuclear mass of the detector material with $Z$ protons and $N=A-Z$ neutrons. In Eq.~(\ref{eq:xsec-cevns}), we take
into account both the vector $\mathcal{Q}_V$ and axial vector $\mathcal{Q}_A$ contributions~\cite{Kosmas:2015vsa} 
\begin{equation}
\begin{aligned}
\mathcal{Q}_V =&  \left[ 2(g_{u}^{L} + g_{u}^{R}) + (g_{d}^{L} + g_{d}^{R}) \right] Z  + \left[ (g_{u}^{L} + g_{u}^{R}) +2(g_{d}^{L} + g_{d}^{R}) \right] N   \, , \\
\mathcal{Q}_A =&  \left[ 2(g_{u}^{L} - g_{u}^{R}) + (g_{d}^{L} - g_{d}^{R}) \right] (\delta Z) + \left[ (g_{u}^{L} - g_{u}^{R}) +2(g_{d}^{L} - g_{d}^{R}) \right] (\delta N)   \, ,
\end{aligned}
\end{equation}
with the abbreviations $(\delta Z)=Z_+ - Z_-$ and $(\delta N)=N_+ - N_-$, where $Z_+ \, (N_+)$ and $Z_- \, (N_-)$ refers to total number
of protons (neutrons) with spin up or down~\cite{Barranco:2005yy}. The left- and right-handed couplings of $u$ and $d$ quarks to the
$Z$-boson including radiative corrections~\cite{Tanabashi:2018oca} are written in terms of the weak mixing-angle 
$\hat{s}^2_Z \equiv \sin^2 \theta_W= 0.23120$ as
\begin{equation}
\begin{aligned}
g_{u}^{L} =& \rho_{\nu N}^{NC} \left( \frac{1}{2}-\frac{2}{3} \hat{\kappa}_{\nu N} \hat{s}^2_Z \right) + \lambda^{u,L} \, ,\\
g_{d}^{L} =& \rho_{\nu N}^{NC} \left( -\frac{1}{2}+\frac{1}{3} \hat{\kappa}_{\nu N} \hat{s}^2_Z \right) + \lambda^{d,L} \, ,\\
g_{u}^{R} =& \rho_{\nu N}^{NC} \left(-\frac{2}{3} \hat{\kappa}_{\nu N} \hat{s}^2_Z \right) + \lambda^{u,R} \, ,\\
g_{d}^{R} =& \rho_{\nu N}^{NC} \left(\frac{1}{3} \hat{\kappa}_{\nu N} \hat{s}^2_Z \right) + \lambda^{d,R} \, ,
\end{aligned}
\end{equation}
with $\rho_{\nu N}^{NC} = 1.0082$, $\hat{\kappa}_{\nu N} = 0.9972$, $\lambda^{u,L} = -0.0031$, $\lambda^{d,L} = -0.0025$ and
$\lambda^{d,R} =2\lambda^{u,R} = 3.7 \times 10^{-5}$.
Nuclear form factors are expected to play a key role in the interpretation of \cevns data (for a recent work see Ref.~\cite{Papoulias:2019lfi}). 
At low-momentum transfer, $-q^{\mu} q_{\mu}=-q^2=Q^{2}=2 m_A T_A$, the finite nuclear size in the \cevns cross section is represented by 
the form factor $F(Q^2)$ correction for which we adopt the symmetrized Fermi (SF) approximation~\cite{Sprung_1997} 
\begin{equation}
F \left( Q ^ { 2 } \right) =  \frac { 3 } { Q c \left[ ( Q c ) ^ { 2 } + ( \pi Q a ) ^ { 2 } \right] }  \left[ \frac { \pi Q a } { \sinh ( \pi Q a ) } \right]   \left[ \frac { \pi Q a \sin ( Q c ) } { \tanh ( \pi Q a ) } - Q c \cos ( Q c ) \right] \, ,
\end{equation}
with 
\begin{equation}
c = 1.23 A^{1/3} - 0.60 \, \text{(fm)}, \quad a=0.52 \, \text{(fm)} \, .
\label{eq:SF-vals}
\end{equation}
where $c$ stands for the half density radius and $a$ denotes the diffuseness.

Next, we will calculate the  \cevns cross section in the presence of non-standard electromagnetic neutrino properties. In general, it is expected that the neutrino magnetic moment will only give a subdominant contribution to the \cevns rate~\cite{Vogel:1989iv}.
For sub-keV threshold experiments, however, the contribution of the electromagnetic (EM) \cevns vertex can be dominant~\cite{Kosmas:2015sqa} and may lead to
detectable distortions of the recoil spectrum. The contribution to the \cevns cross section reads 
\begin{equation}
\left( \frac{d \sigma}{dT_A} \right)_{\mathrm{EM}}=\frac{\pi a^2_{\text{EM}} \mu_{\nu}^{2}\,Z^{2}}{m_{e}^{2}}\left(\frac{1-T_A/E_{\nu}}{T_A}\right) F^{2}(Q^{2})\,.
\label{NMM-cross section}
\end{equation}
In this framework, the helicity preserving standard weak interaction cross section (SM) adds incoherently with the helicity-violating EM cross section, so the total cross section is written as 
\begin{equation}
\left( \frac{d \sigma}{dT_A}\right)_{\mathrm{tot}} = \left( \frac{d \sigma}{dT_A} \right)_{\mathrm{SM}} 
+ \left( \frac{d \sigma}{dT_A} 
\right)_{\mathrm{EM}}\, .
\end{equation}
In what follows, we adopt the theoretical expressions for the effective neutrino magnetic moment $\mu_\nu$ parameter in the mass basis, 
derived in Sect.~\ref{sec:theoretical-framework} in order to  constrain the TMM parameters.

\section{Experimental setup}
\label{sec:experimental-setup}

\begin{table*}[t]
\resizebox{\textwidth}{!}{\begin{tabular}{@{}lccccccc@{}}
\toprule
\textbf{Experiment}  & \textbf{detector} & \textbf{mass} & \textbf{threshold} & \textbf{efficiency}& \textbf{exposure} & \textbf{baseline (m)} \\ \midrule
 
\multicolumn{7}{c}{\textbf{SNS}}\\

COHERENT~\cite{Akimov:2017ade}  & CsI[Na]  & 14.57~kg [100~kg]  &  5~keV [1~keV]   & Eq.~(\ref{eq:eff-SNS}) [100\%]  & 308.1~days [10~yr]  & 19.3  \\
COHERENT~\cite{Akimov:2018ghi}  & HPGe     & 15~kg    [100~kg]  &  5~keV [1~keV]   & 50\%                  [100\%]  &        308.1~days [10~yr]  & 22    \\
COHERENT~\cite{Akimov:2018ghi}  & LAr      & 1~ton    [10~ton]  & 20~keV [10~keV]  & 50\%                  [100\%]  &         308.1~days [10~yr]  & 29    \\
COHERENT~\cite{Akimov:2018ghi}  & NaI[Tl]  & 2~ton    [10~ton]  & 13~keV [5~keV]   & 50\%                  [100\%]  &        308.1~days [10~yr]  & 28    \\
                                
\multicolumn{7}{c}{\textbf{Reactor}}\\

CONUS~\cite{conus}                      & Ge         & 3.85~kg  [100~kg]  & 100 eV    & 50\% [100\%]  & 1~yr [10~yr] & 17   \\
CONNIE~\cite{Aguilar-Arevalo:2016qen}   & Si         & 1~kg     [100~kg]  & ~28 eV    & 50\% [100\%]  & 1~yr [10~yr] & 30   \\
MINER~\cite{Agnolet:2016zir}            & 2Ge:1Si    & 1~kg     [100~kg]  & 100 eV    & 50\% [100\%]  & 1~yr [10~yr] & ~2   \\
TEXONO~\cite{Wong:2010zzc}              & Ge         & 1~kg     [100~kg]  & 100 eV    & 50\% [100\%]  & 1~yr [10~yr] & 28   \\
RED100~\cite{Akimov:2016xdc}            & Xe         & 100~kg   [100~kg]  & 500 eV    & 50\% [100\%]  & 1~yr [10~yr] & 19   \\ 
\bottomrule 
\end{tabular}}
\caption{ CE$\nu$NS experimental setups considered in the present study. 
 Values corresponding to the future setups are given in square brackets.} 
\label{table:exper}
\end{table*}
%

\begin{figure}[t]
\includegraphics[width=0.5\textwidth]{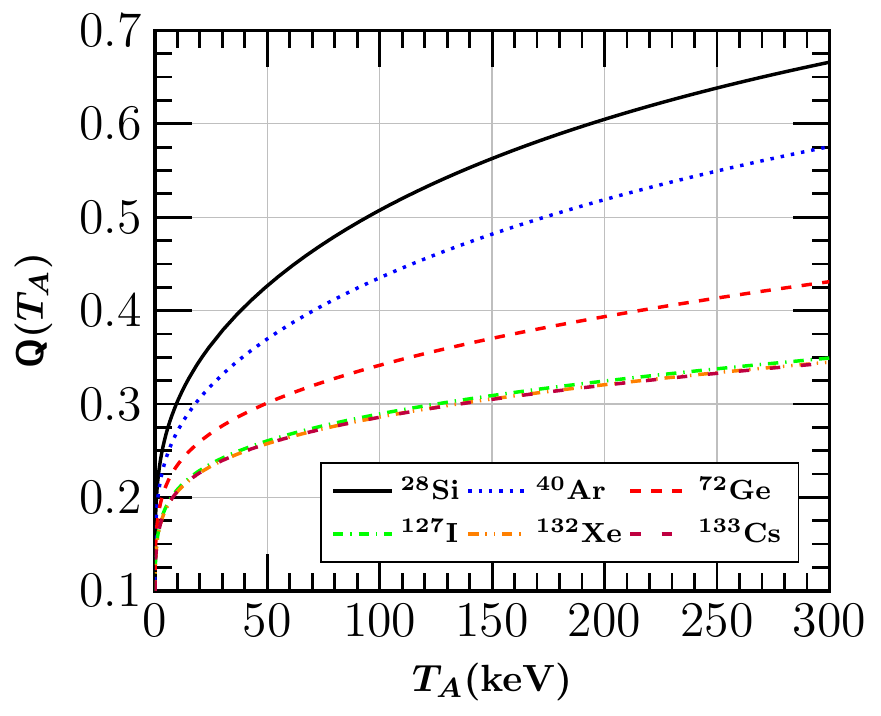}
\caption{Quenching factor with respect to the nuclear recoil energy $T_{A}$ for the detector nuclei of the CE$\nu$NS experiments (see Table~\ref{table:exper}).}
\label{fig:quenching}
\end{figure}

We find it useful to devote a separate section to discuss the main features of our calculation procedure. For ionization detectors, a
significant part of the nuclear recoil energy is lost into heat, so the measured energy (electron equivalent energy) is lower. We take
into account this energy loss by considering the quenching factor $\mathsf{Q}(T_A)$, that is calculated from the Lindhard theory~\cite{Lewin:1995rx} 
\begin{equation}
\mathsf{Q}(T_A) = \frac{\kappa g(\gamma)}{1+\kappa g(\gamma)}\, ,
\end{equation}
with $g(\gamma)= 3 \gamma^{0.15} + 0.7 \gamma^{0.6}+\gamma$ and $\gamma  = 11.5 \, T_A(\text{keV}) Z^{-7/3}$, $\kappa = 0.133 \, Z^{2/3} A^{-1/2}$. 
Figure~\ref{fig:quenching} presents the effect of the quenching factor with respect to the nuclear recoil energy $T_A$ for all nuclei used in this work. 

Here, we not only examine the sensitivity of \cevns experiments to TMMs according to their current setup, but also explore their long-term prospects. 
To this purpose, we also consider a future experimental setup with larger detector mass, 
improved threshold capabilities and an increased time of exposure. Note that, even in the adopted future setups, the input values follow the
proposal of each experiment. Therefore, they are quite realistic, leading to reasonable projected sensitivities. The details of the
assumed experimental setups are shown in Table~\ref{table:exper}. 

\subsection{Reactor Neutrinos}

In reactor neutrino experiments, electron antineutrinos are generated by the beta-decay of the fission products of $^{235}$U, $^{238}$U,
$^{239}$Pu and $^{241}$Pu. We calculate the energy distribution $f_{\bar{\nu}_e}(E_\nu)$ by employing the expansion of Ref.~\cite{Mention:2011rk}, 
whereas for energies below 2~MeV, due to lack of experimental data, we consider the theoretical calculations given in Ref.~\cite{Kopeikin:1997ve}. 
The neutrino flux $\Phi_\nu$ depends on the power of the reactor plant and the baseline for the relevant experiment (see 
Refs.~\cite{Wong:2010zzc,Aguilar-Arevalo:2016qen,Agnolet:2016zir,conus,Akimov:2016xdc}). In all cases we assume a flat detector 
efficiency  of 50\% in the event identification and a benchmark of 1~yr data taking. 

%

\subsection{Neutrinos at the Spallation Neutron Source}

The first \cevns measurement by COHERENT became feasible by employing a CsI[Na] detector with mass $m_{\mathrm{det}}=14.57$~kg located at a
baseline of $L=19.3$~m from the DAR-$\pi$ source with an exposure time of 308.1~days. Following the recipe of the COHERENT data release~\cite{Akimov:2018vzs}, 
we adequately simulate the DAR-$\pi$ neutrino spectra in terms of the pion  and muon  masses, $m_\pi$ and $m_\mu$, following the Michel spectrum~\cite{Louis:2009zza} 
\begin{equation}
\begin{aligned} 
f_{\nu_\mu}(E_\nu) & =  \delta\left(E_\nu-\frac{m_{\pi}^{2}-m_{\mu}^{2}}{2 m_{\pi}}\right) \, , \\ 
f_{\bar{\nu}_\mu}(E_\nu) & = \frac{64 E^{2}_\nu}{m_{\mu}^{3}}\left(\frac{3}{4}-\frac{E_\nu}{m_{\mu}}\right) \, ,\\ 
f_{\nu_e}(E_\nu) & = \frac{192 E^{2}_\nu}{m_{\mu}^{3}}\left(\frac{1}{2}-\frac{E_\nu}{m_{\mu}}\right) \, ,
\end{aligned}
\label{labor-nu}
\end{equation}
where $E_{\nu}^{\text{max}} \leq m_{\mu}/2 \approx 52.8$~MeV. The latter accounts for the monochromatic muon neutrino  beam
($E_\nu=29.9$ MeV) produced from pion decay at rest, $\pi^+ \to \mu^{+} \nu_{\mu} $ (prompt flux with $\tau=26 \, \mathrm{ns}$), and
the subsequent $\nu_{e}$ and ${\bar{\nu}_{\mu}}$ neutrino beams resulting from muon decay $\mu^{+} \to \nu_{e} e^{+}
\bar{\nu}_{\mu}$ (delayed flux with $\tau=2.2 \, \mathrm{\mu  s}$)~\cite{Efremenko:2008an}.  The neutrino flux is $\Phi_{\nu} = r
N_{\mathrm{POT}}/4 \pi L^2$, with $r=0.08$ representing the number of neutrinos per flavor produced for each proton on target (POT), e.g.
$N_{\mathrm{POT}}=1.76 \times 10^{23}$ corresponding to the 308.1~days of exposure during the first run. For the future COHERENT detector
subsystems HPGe, LAr and NaI[Tl], we assume an exposure period of 1~yr, which corresponds to $N_{\mathrm{POT}}=2.09 \times 10^{23}$. 

The COHERENT signal was detected through photoelectron (PE) measurements, hence, in our simulations we translate the energy of the
scattered nucleus $T_A$ in terms of the number of the observed PE, $n_{\text{PE}}$, through the relation~\cite{Akimov:2017ade} 
\begin{equation}
n_{\mathrm{PE}} = 1.17 \frac{T_A}{(\mathrm{keV})}\, ,
\end{equation}
taking also into consideration the photoelectron dependence of the detector efficiency $\mathcal{A}(x)$, required for determining
the expected event rate below and given by~\cite{Akimov:2018vzs}   
\begin{equation}
\mathcal{A}(x) = \frac{k_1}{1 + e^{- k_2 \left(x - x_0 \right)}} \Theta(x) \, ,
\label{eq:eff-SNS}
\end{equation}
with $k_1= 0.6655$, $k_2= 0.4942$, $x_0= 10.8507$ and the Heaviside function 
\begin{equation}
\Theta(x) = \left\{ \begin{array}{ll}{0} & {x < 5} \\ {0.5} & {5 \leq x < 6} \\ {1} & {x \geq 6 \, .} \end{array} \right.
\end{equation}
As in the case of reactor experiments, due to the lack of relevant information for the next generation detector subsystems 
HPGe, LAr and NaI[Tl], we assume a flat efficiency of $\mathcal{A}(T_A)=0.5$.

\section{Numerical Results}
\label{sec:results}

For a given \cevns experiment, the total cross section is evaluated as a sum of the individual cross sections corresponding to each isotope
composing the detector material. By taking into account the stoichiometric ratio of the atom, $\eta$, and the detector mass,
$m_\text{det}$, the number of target nuclei per isotope is evaluated through  Avogadro's number, $N_A$ 
\begin{equation}
N_{\mathrm{targ}}^x = \frac{m_{\mathrm{det}} \eta_x}{\sum_x A_x \eta_x}  N_A \, ,
\end{equation}
while the total number of events expected above threshold $T_{\mathrm{th}}$ (see Table~\ref{table:exper}) reads~\cite{Kosmas:2017tsq} 
\begin{equation}
\begin{aligned}
N_{\mathrm{theor}} = \sum_{\nu_\alpha} \sum_{x = \mathrm{isotope}} \mathcal{F}_x   \int_{T_{\mathrm{th}}}^{T_{A}^{\mathrm{max}}} \int_{E_{\nu}^{\mathrm{min}}}^{E_{\nu}^{\mathrm{max}}} f_{\nu_{\alpha}}(E_\nu) \mathcal{ A } ( T_A ) \left(\frac{{d \sigma}_{x}}{dT_A}(E_\nu, T_A) \right)_{\mathrm{tot}}   dE_\nu dT_A \, ,
\end{aligned}
\label{eq:events}
\end{equation}
where the luminosity for a detector with target material $x$ is given by $\mathcal{F}_x = N_{\mathrm{targ}}^x \Phi_\nu$ 
and $E_{\nu}^{\mathrm{min}}= \sqrt{m_A T_A /2}$ is the minimum incident neutrino energy to produce a nuclear recoil. 
Notice that we sum over all possible incident neutrino flavors $\alpha$ scattering off a detector with all possible isotopes $x$.
It is worth mentioning that potential contributions to the event rate from detector dopants are safely ignored, since they are of the order $10^{-5}$ -- $10^{-4}$~\cite{Collar:2014lya}.

\begin{figure}[t]
\includegraphics[width=\textwidth]{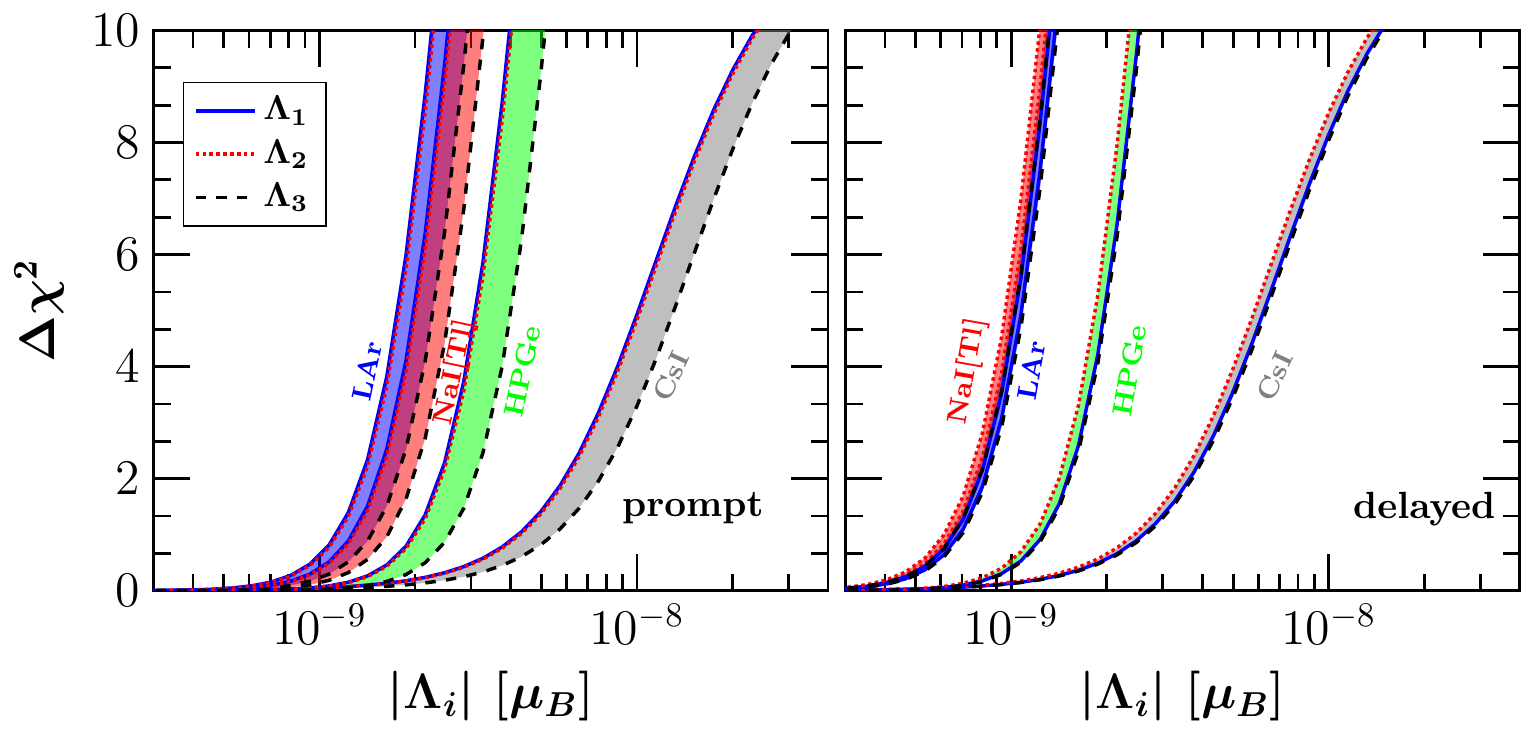}
\includegraphics[width=\textwidth]{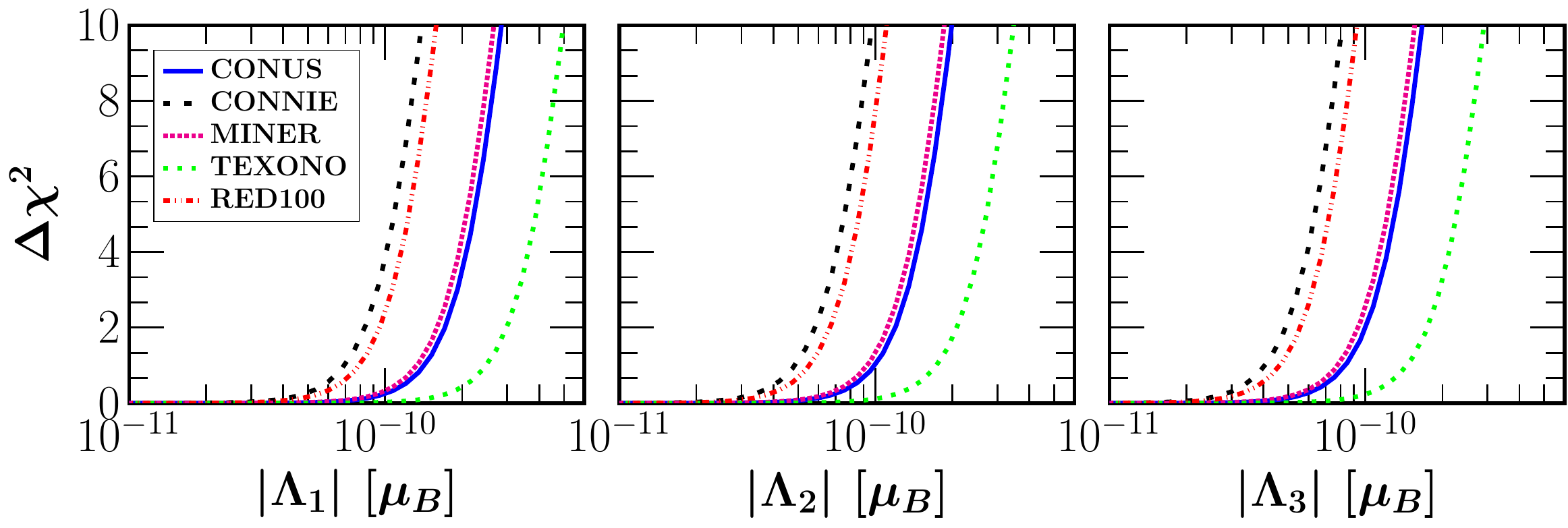}
\caption{$\Delta \chi^2$ profiles for every element of the TMM matrix, $\left \vert \Lambda_i \right \vert$, at \cevns experiments 
for vanishing $\left \vert \Lambda_j \right \vert$, $\left \vert \Lambda_k \right \vert$ and all phases set to zero. The results in 
the upper (lower) panel are for SNS (reactor) neutrino experiments in their current setup. The color bands in the upper panels 
indicate the limits expected from each SNS experiment.} 
\label{fig:Li_1param}
\end{figure}
%

In order to extract the current constraints on  TMMs $\left \vert \Lambda_i \right \vert$ from the first phase of COHERENT (with a
CsI detector), we perform a  statistical analysis using the following $\chi^2$ function~\cite{Akimov:2017ade} 
\begin{equation}
\begin{aligned}
\chi^2 (\mathcal{S}) =  \underset{\mathtt{a}_1, \mathtt{a}_2}{\mathrm{min}} \Bigg [ \frac{\left(N_{\mathrm{meas}} - N_{\mathrm{theor}}(\mathcal{S}) [1+\mathtt{a}_1] - B_{0n} [1+\mathtt{a}_2] \right)^2}{(\sigma_{\mathrm{stat}})^2} 
  + \left(\frac{\mathtt{a}_1}{\sigma_{\mathtt{a}_1}} \right)^2 + \left(\frac{\mathtt{a}_2}{\sigma_{\mathtt{a}_2}} \right)^2 \Bigg ] \, .
\end{aligned}
\label{eq:chi}
\end{equation}
Here, the measured number of events is $N_{\mathrm{meas}}=142$, while $\mathtt{a}_1$ and $\mathtt{a}_2$ are the systematic parameters 
accounting for the uncertainties on the signal and background rates, respectively, with fractional uncertainties $\sigma_{\mathtt{a}_1} =
0.28$ and $\sigma_{\mathtt{a}_2} = 0.25$. Following Ref.~\cite{Akimov:2017ade}, the statistical uncertainty is given by
$\sigma_{\mathrm{stat}}=\sqrt{N_{\mathrm{meas}} + B_{0n} + 2 B_{ss}}$, where the quantities $B_{0n}=6$ and $B_{ss}=405$ denote the beam-on
prompt neutron and the steady-state background events respectively. Our statistical analysis regarding  reactor as well as the
next generation of COHERENT \cevns experiments, within the framework of  current and future setups, is based on a single
nuisance parameter. In this case, the $\chi^2$ function is defined as 
\begin{equation}
\chi^2(\mathcal{S}) = \underset{\mathtt{a}}{\mathrm{min}} \Bigg [ \frac{\left(N_{\text{meas}} - N_{\text{theor}}(\mathcal{S}) [1+\mathtt{a}]\right)^2} {(1 + \sigma_{\text{stat}}) N_{\text{meas}}}   + \left( \frac{\mathtt{a}}{\sigma_{\text{sys}}}\right)^2 \Bigg] \, ,
\end{equation}
where we adopt the values $\sigma_{stat}= \sigma_{\text{sys}} = 0.2$ $(0.1)$ for the current (future) setups. In order to probe TMMs, in what follows we perform a 
minimization over the nuisance parameter $\mathtt{a}$ and calculate $\Delta \chi^2(\mathcal{S})= \chi^2(\mathcal{S}) - \chi^2_{\mathrm{min}}(\mathcal{S})$, with $\mathcal{S}\equiv \{\left \vert \Lambda_i \right \vert, \xi_i, \delta_{\text{CP}}\}$ denoting the  set of parameters entering the 
definition of the effective neutrino magnetic moment.

%
\begin{figure*}[t]
\includegraphics[width=\textwidth]{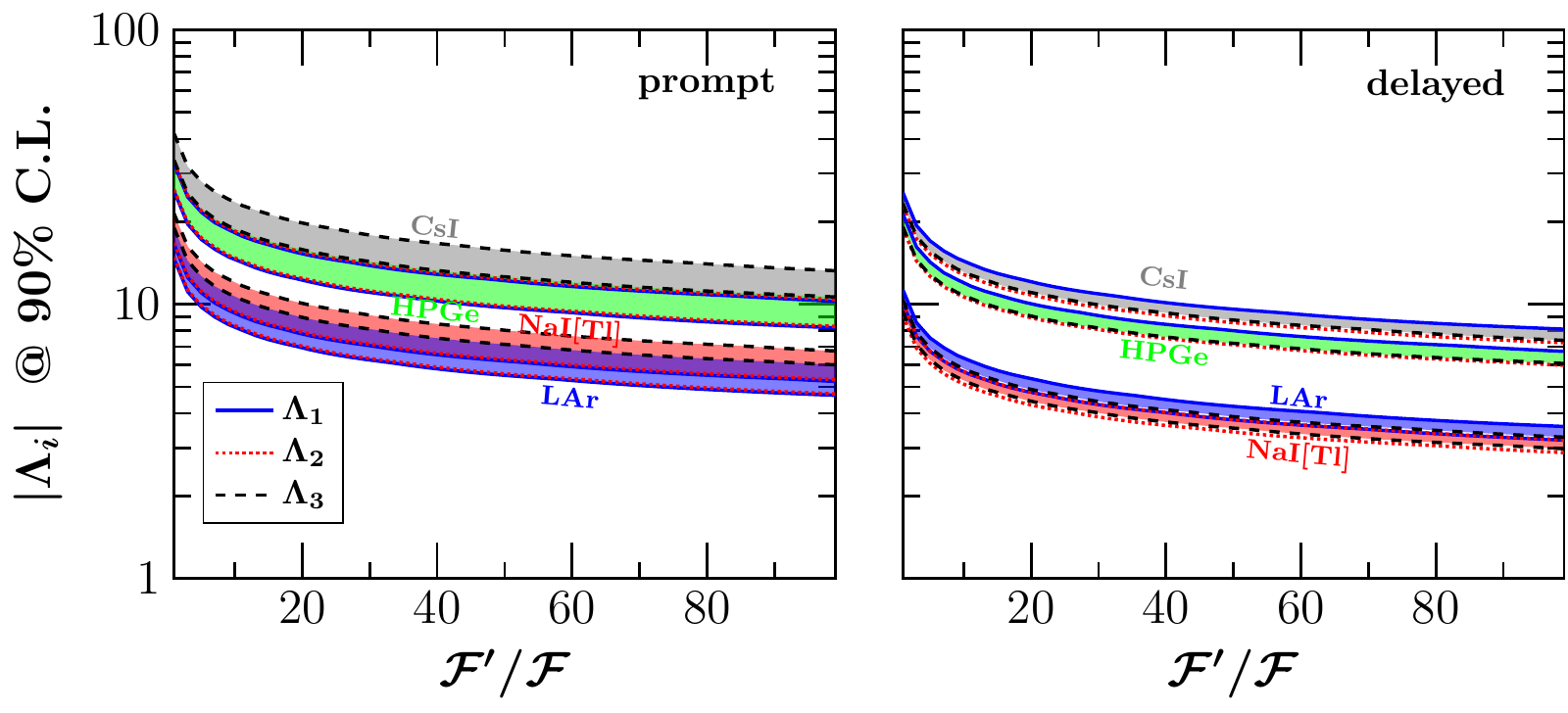}
\includegraphics[width=\textwidth]{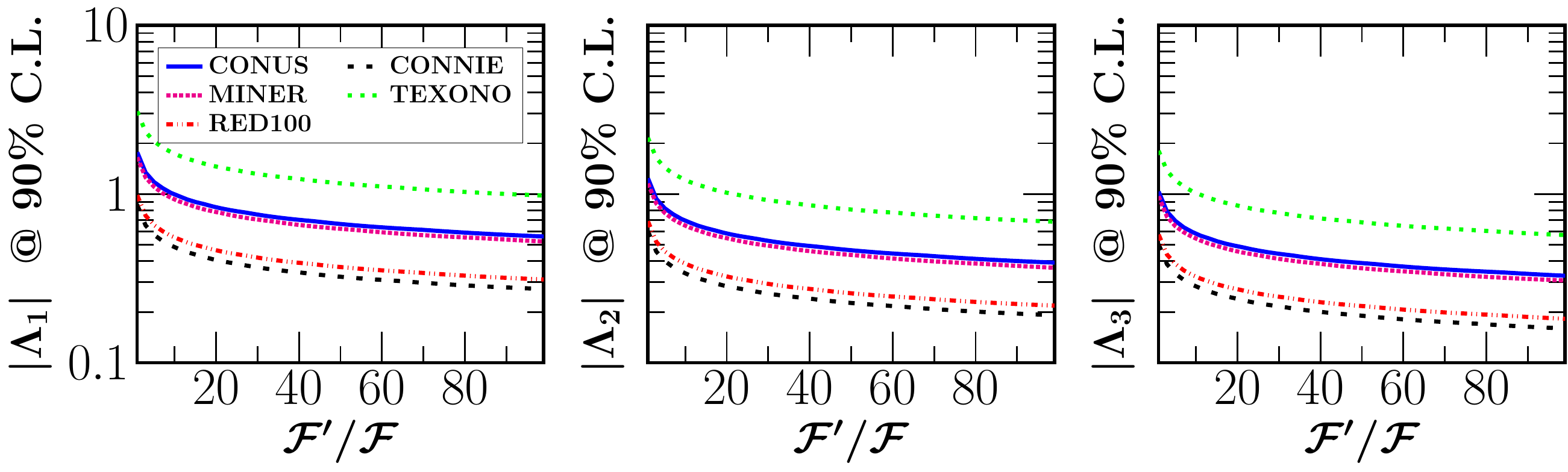}
\caption{Variation of the  90\% C.L. limits on $\left \vert \Lambda_i \right \vert$ as a function of the total luminosity 
$\mathcal{F^\prime}$ for SNS (upper panel) and reactor (lower panel) neutrino experiments. In all cases we assume vanishing 
$\left \vert \Lambda_j \right \vert$, $\left \vert \Lambda_k \right \vert$ and all phases set to zero. The results are shown in 
units $10^{-10} \mu_B$.}
\label{fig:Li_1param_NPOT}
\end{figure*}

We begin our sensitivity analysis by considering a single non-vanishing TMM parameter $ \left \vert \Lambda_i \right \vert$ at a
time, in the current setup. As a first step, for the sake of simplicity, in our calculations we set all complex phases to zero, 
assuming all TMMs as real parameters. We will discuss the impact of non-zero  phases on our reported sensitivities at the end of 
this section. 
The extracted constraints from the first \cevns measurement in CsI along with the projected sensitivities from the next phase HPGe, LAr and NaI[Tl] COHERENT 
subsystems, are shown in the upper panel of Fig.~\ref{fig:Li_1param} for prompt and delayed neutrinos. From the first run of the COHERENT experiment, the following 90\% C.L. bounds are obtained from the prompt (delayed) neutrino beams 
\begin{equation}
\begin{aligned}
\left \vert \Lambda_1 \right \vert < \,& 69.2 \, \,  (54.5) \, \times 10^{-10}\, \mu_B \, ,\\
\left \vert \Lambda_2 \right \vert <  \, & 70.2 \, \, (48.7) \, \times 10^{-10}\, \mu_B \, ,\\
\left \vert \Lambda_2 \right \vert < \, & 89.6 \, \,  (49.8) \,  \times 10^{-10}\, \mu_B \, .
\end{aligned}
\end{equation}
%
%
\begin{figure*}[t]
\includegraphics[width=\linewidth]{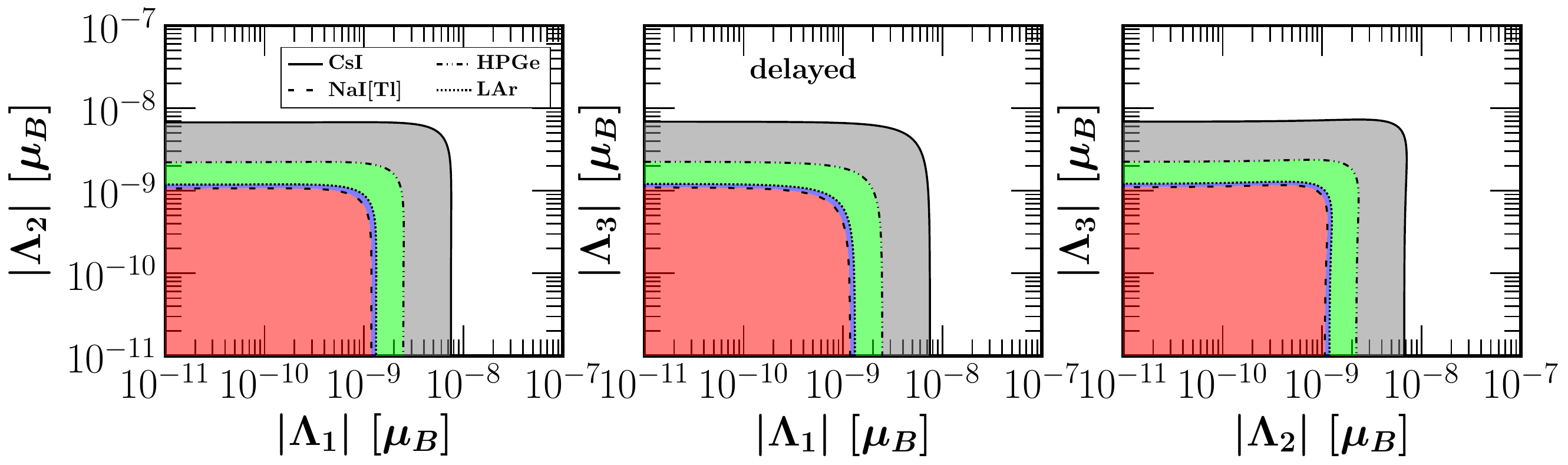}
\includegraphics[width=\linewidth]{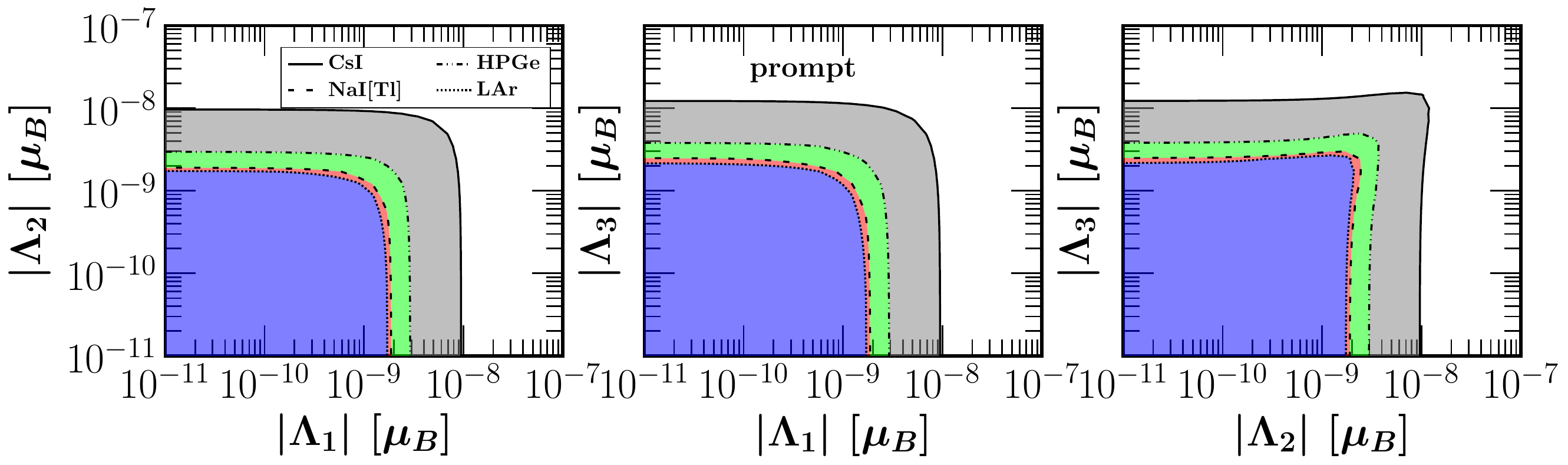}
\caption{Regions in the $\left \vert \Lambda_i \right \vert - \left \vert \Lambda_j \right \vert$ parameter space allowed 
at 90\% C.L. by current data of the COHERENT experiment (CsI detector) for vanishing values of the undisplayed 
$\left \vert \Lambda_k \right \vert$ and all phases.
The upper (lower) panel presents the results for delayed (prompt) neutrinos in the current setup. 
We also display the projected sensitivities for the HPGe, LAr and NaI[Tl] detector subsystems of COHERENT. 
The color labeling is same as in Fig.~\ref{fig:Li_1param}.} 
\label{fig:Li_vs_Lj_SNS}
\end{figure*}
%
\begin{figure*}[t]
\includegraphics[width=\linewidth]{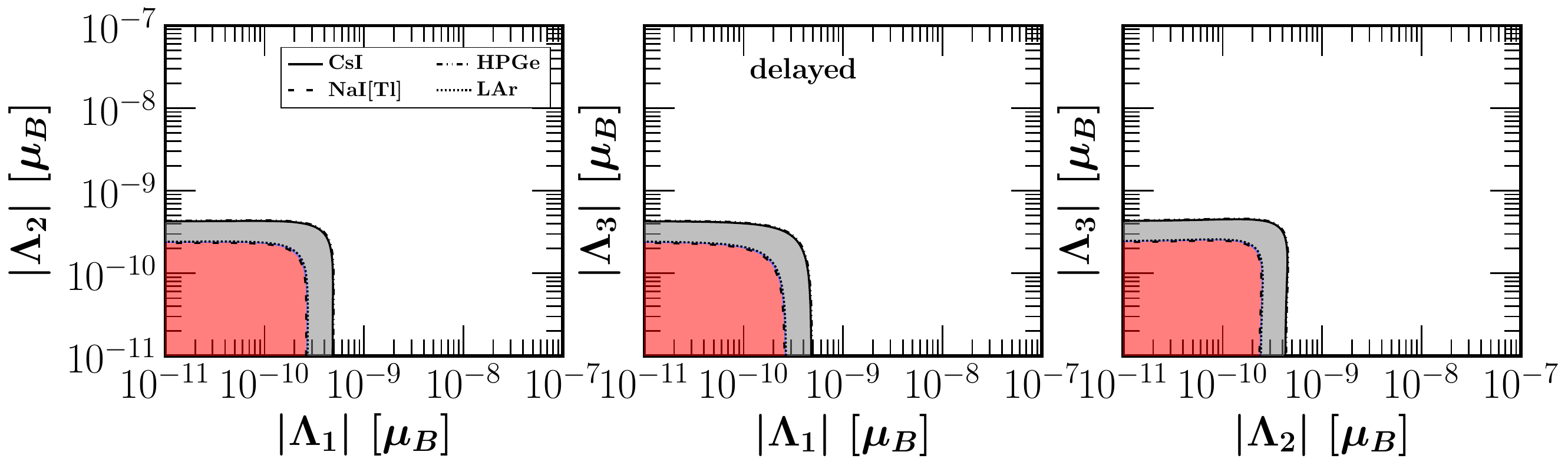}
\includegraphics[width=\linewidth]{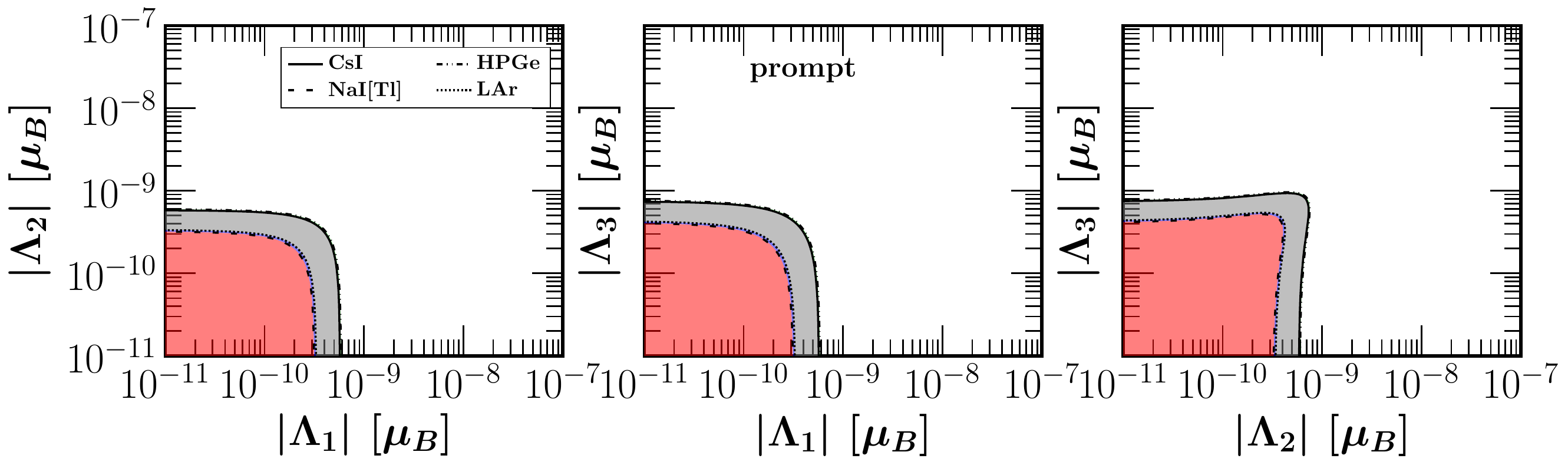}
\caption{Same as Fig.~\ref{fig:Li_vs_Lj_SNS}  for the case of the future experimental setup at the SNS.}
\label{fig:Li_vs_Lj_SNS_future}
\end{figure*}
%
\begin{figure*}[t]
\includegraphics[width=\textwidth]{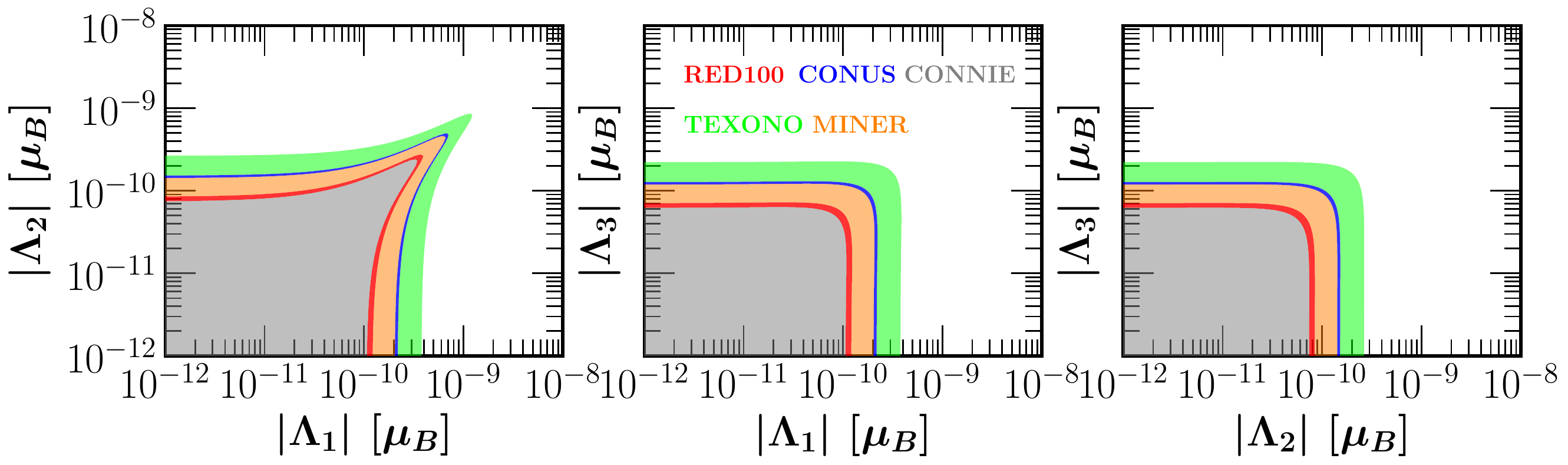}
\includegraphics[width=\textwidth]{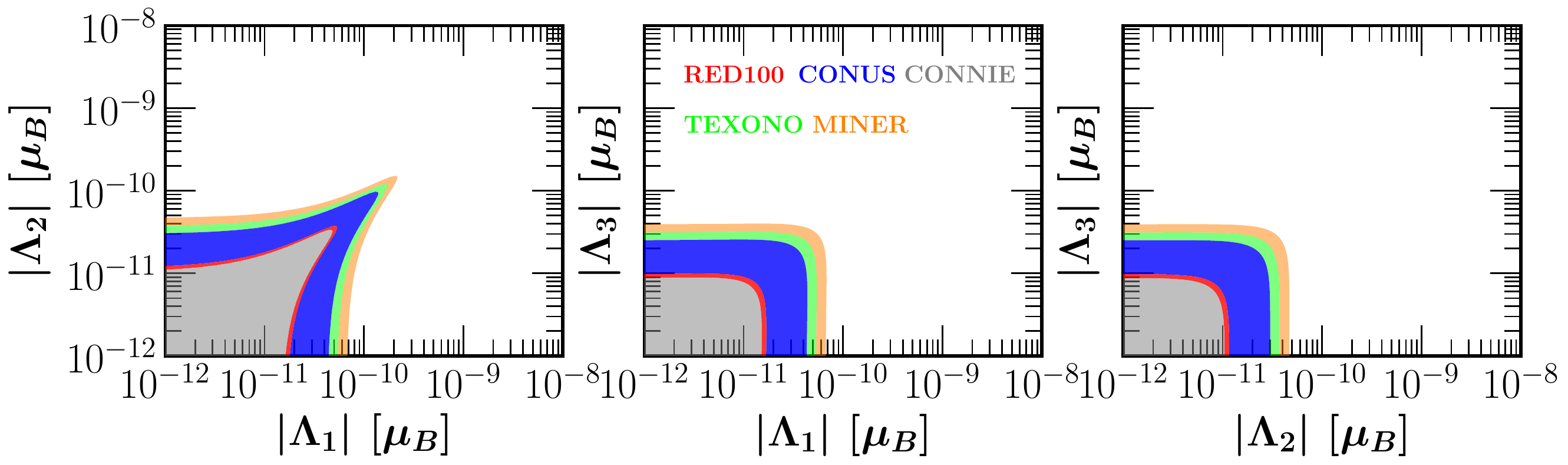}
\caption{Projected 90\% C.L. sensitivities in the $\left \vert \Lambda_i \right \vert - \left \vert \Lambda_j \right \vert$ plane 
assuming vanishing values of the undisplayed $\left \vert \Lambda_k \right \vert$ and CP phases. Upper and lower panels correspond
to the current and future configurations of reactor neutrino experiments.}
\label{fig:Li_vs_Lj_reactors}
\end{figure*}
%

On the other hand, the lower panel of Fig.~\ref{fig:Li_1param} presents the corresponding 
projected sensitivity from the reactor \cevns experiments CONUS, CONNIE, MINER, TEXONO and RED100.  
The results presented in Fig.~\ref{fig:Li_1param} indicate that the prospects for probing electromagnetic neutrino properties are better for reactor-based experiments.
 This is a direct consequence of their 
sub-keV recoil threshold capabilities in conjunction with the fact that the reactor neutrino energy distribution is peaked at much lower energies, compared to 
DAR-$\pi$ neutrinos.  We stress, however, that when considering the full SNS beam, instead of the individual prompt and delayed components, this difference is 
reduced significantly.
As an illustrative example, by  assuming the full SNS beam in the current configuration of the CsI detector, the corresponding 90\% C.L. upper bounds on 
($\left \vert \Lambda_1 \right \vert$, $\left \vert \Lambda_2 \right \vert$, $\left \vert \Lambda_3 \right \vert$) are  (42.8, 40.0, 43.6) in units of 
$10^{-10}\, \mu_B$. Similarly, for the future detector materials of COHERENT, the projected sensitivities read,  Ge: (16.5, 15.3, 16.6), LAr: (8.9, 8.4, 9.1) and 
NaI: (8.6, 8.0, 8.6), all in units of $10^{-10}\, \mu_B$.  

For completeness, we now examine the attainable sensitivity for different values of the factor $\mathcal{F}$ which 
corresponds to the luminosity of each studied experiment, entering in the calculation of the event number in Eq.~(\ref{eq:events}).
To be conservative, we fix all other inputs  to their default values according to the current setup and, as previously, 
we assume all TMMs to be real. We then calculate the sensitivities on $\left \vert \Lambda_i \right \vert$ for SNS
and reactor \cevns experiments by scaling-up our simulations with the new luminosity factor $\mathcal{F}^\prime$. 
With this approach, it becomes feasible to probe the sensitivity on TMMs for several combinations of detector mass, exposure 
period, detector baselines and power of the source. To motivate this approach, we recall that the SNS should
increase its operation power and also that MINER is planning towards a moveable core strategy. Our
corresponding results are depicted in Fig.~\ref{fig:Li_1param_NPOT}. They show that there is a significant 
improvement by adopting scale-up factors of the order of $\mathcal{F}^\prime / \mathcal{F}
\lesssim 40$, whereas beyond that point the improvement becomes weaker. 

%
\begin{figure}[t]
\includegraphics[width=\textwidth]{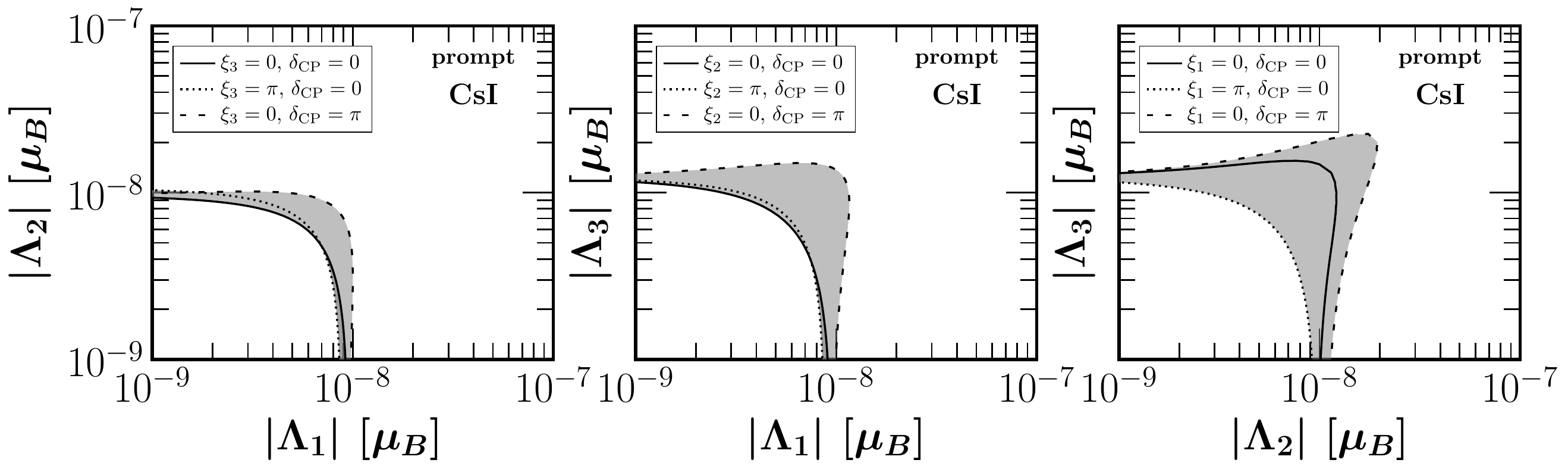}
\includegraphics[width=\textwidth]{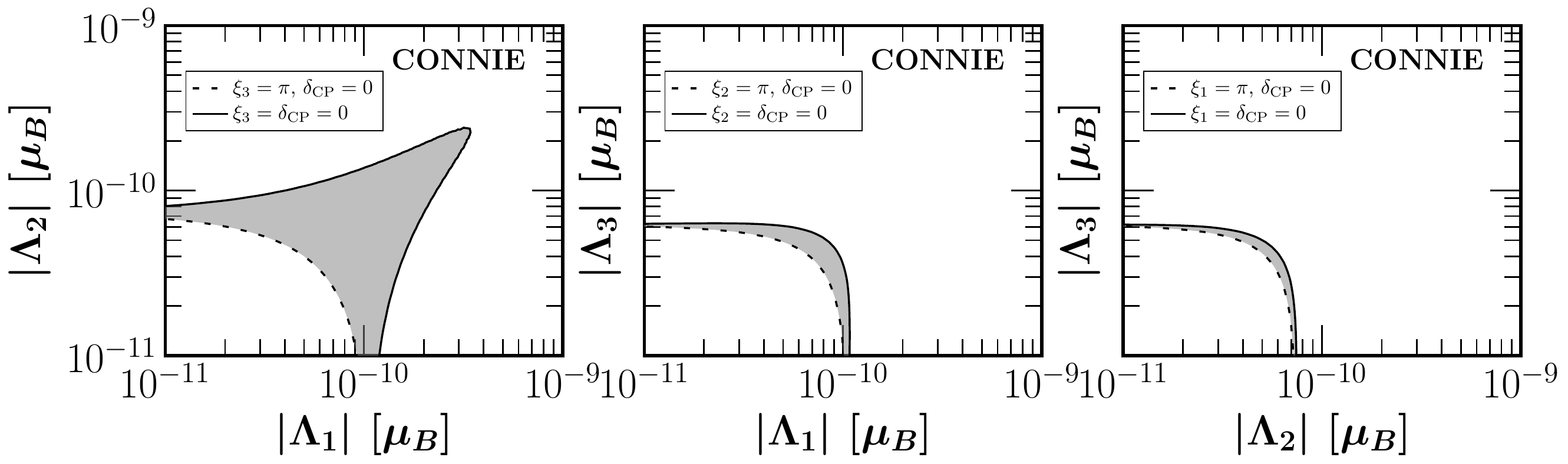}
\caption{Projected 90\% C.L. contours in the $\left \vert \Lambda_i \right \vert-\left \vert \Lambda_j \right \vert$ 
plane from the analysis of the prompt beam at the SNS (upper panel) and from the reactor neutrino experiment CONNIE (lower panel), 
for different values of the Majorana CP violating phases. As before, we have assumed a vanishing value for the undisplayed 
$\left \vert \Lambda_k \right \vert$.}
\label{fig:phases_SNS}
\end{figure}
%

We are now interested in exploring simultaneous constraints on the effective neutrino magnetic moment parameters from the current 
as well as projected \cevns data, according to the setups reported in Table~\ref{table:exper}. Assuming two real non-vanishing TMMs 
at a time, in Fig.~\ref{fig:Li_vs_Lj_SNS} we present the allowed regions in the 
$\left \vert \Lambda_i \right \vert - \left \vert \Lambda_j \right \vert$ plane extracted from the available CsI COHERENT data for the prompt and delayed beams. 
The corresponding regions for the next generation of COHERENT detectors are also shown. As can be seen from the plot, the 
installation of LAr and NaI[Tl] detector subsystems will  offer improvements of about one order of magnitude, after one year 
of data taking. Figure~\ref{fig:Li_vs_Lj_SNS_future} shows the projected sensitivities expected at various detector subsystems 
of COHERENT in the future setup, with an improvement of at least one order of magnitude. 
As commented above, a combined analysis of the full SNS beam, would have the potential to place even 
stronger limits. Turning now to reactor-based \cevns experiments, we perform a similar analysis as previously described  and present the projected sensitivities assuming 
the current (future) setup in the upper (lower) panel of Fig.~\ref{fig:Li_vs_Lj_reactors}. In both cases, the two-dimensional contour plots confirm that \cevns 
experiments can be regarded as suitable facilities to probe Majorana electromagnetic properties with improved sensitivity. 
Indeed, as we will see in Sect.~\ref{sec:other-experiments}, with the next-generation upgrades, future measurements will have the potential 
to significantly improve upon the best current constraints, obtained from Borexino solar neutrino data.
Finally, from Figs.~\ref{fig:Li_vs_Lj_SNS}-\ref{fig:Li_vs_Lj_reactors}, one sees that the resulting sensitivities have a slightly different shape in the $\left \vert \Lambda_2 \right \vert - \left \vert \Lambda_3 \right \vert$ plane compared to the other two panels for the case of SNS neutrinos. On the other hand, reactor neutrino experiments show a similar but more pronounced effect in the $\left \vert \Lambda_1 \right \vert - \left \vert \Lambda_2 \right \vert$ plane. This is due to its stronger dependence on the mixing angles and CP phases.

Before closing our discussion concerning the prospects of probing TMMs at \cevns facilities, we examine the robustness of our results by 
exploring the impact of the CP phases on the derived sensitivities in the $\left \vert \Lambda_i \right \vert - \left \vert \Lambda_j \right \vert$ 
plane. As before, we assume a vanishing value for the remaining $\left \vert \Lambda_k \right \vert$. 
As an illustrative example, Fig.~\ref{fig:phases_SNS} shows the different 90\% C.L. contours in the current setup obtained from the prompt beam at the COHERENT experiment (upper panel) and the projected reactor neutrino experiment CONNIE (lower panel). 
For SNS neutrinos, we have verified that the most conservative sensitivity (outer curve) corresponds to $\xi_k=0$ and $ \delta_{\text{CP}}=\pi$, while the strongest one
 (inner curve) corresponds to $\xi_k=\pi$ and $\delta_{\text{CP}}=0$. 
On the other hand, for reactor neutrinos the most conservative sensitivity contour (outer curve) corresponds to $\xi_k=\delta_{\text{CP}}=0$, while the 
most aggressive one (inner curve) is obtained for $\xi_k=\pi$ and $\delta_{\text{CP}}=0$. 
All calculations refer to the current configuration, so that the solid lines correspond to the results presented in Figs.~\ref{fig:Li_vs_Lj_SNS} 
and~\ref{fig:Li_vs_Lj_reactors} assuming real TMMs.


\section{Comparison with the current Borexino limit}
\label{sec:other-experiments}

As already discussed, the neutrino magnetic moment observable at a given experiment is actually an
effective parameter depending on the neutrino mixing parameters as well as the oscillation factor describing the neutrino propagation between the
source and detection points~\cite{Vogel:1989iv,Beacom:1999wx}, i.e.
\begin{equation}
\left(\mu^M_{\nu,\text{eff}}\right)^2 (L, E_\nu) = \sum_j \Big \vert \sum_i U^\ast_{\alpha i} e^{-i\, \Delta m^2_{ij} L /2 E_\nu} \tilde{\lambda}_{ij} \Big \vert^2 \, .
\label{NMM-observable}
\end{equation}
Note that, for the case of the short baseline \cevns experiments discussed in the previous sections, the oscillation factor can be
safely ignored, since there is no time for neutrino oscillations to develop.

\begin{figure*}[t]
\includegraphics[width=\textwidth]{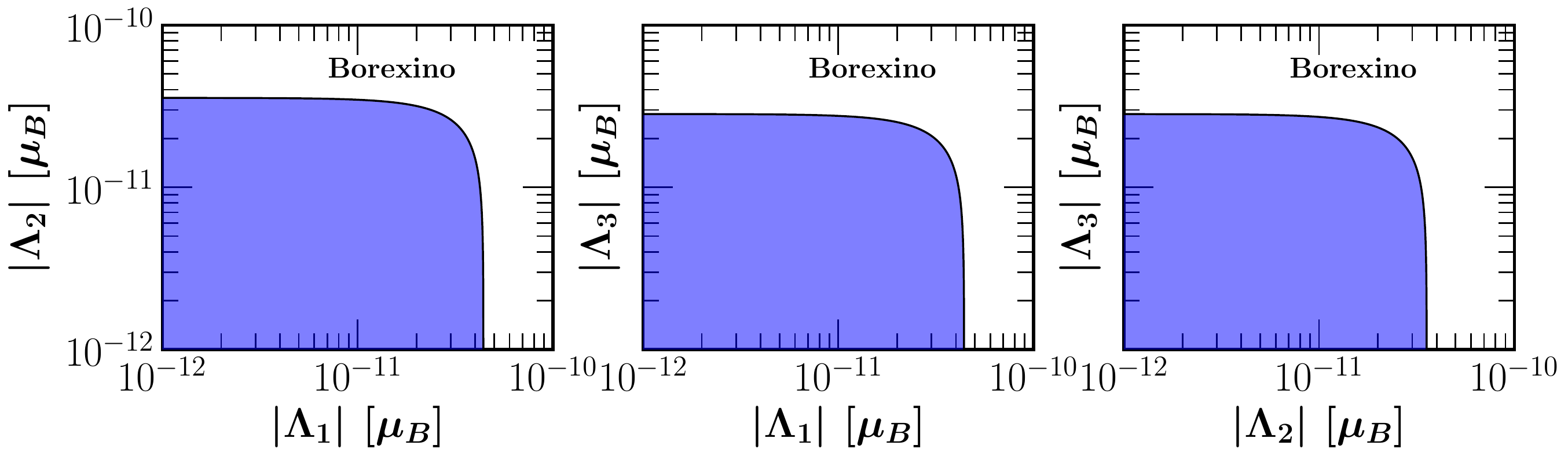}
\caption{
Regions in the $\left \vert \Lambda_i \right \vert - \left \vert \Lambda_j \right \vert$ plane allowed by Borexino solar data at 90\% C. L. As before,
we assume a vanishing value for the undisplayed $\left \vert \Lambda_k \right \vert$.}
\label{fig:Li_vs_Lj_Borexino}
\end{figure*}

To compare our results with current limits on TMMs, we analyze the recent solar neutrino data from 
Borexino phase-II~\cite{Borexino:2017fbd} (see also Refs.~\cite{Guzzo:2012rf,Barranco:2017zeq,Khan:2017djo}). In this case, the expression for the effective neutrino magnetic moment for
solar neutrinos, in the mass basis is given by~\cite{Canas:2015yoa}
\begin{equation}\label{eq:nmm_sun}
(\mu^{M}_{\nu,\,\text{sol}})^{2} = |\mathbf{\Lambda}|^{2} -
  c^{2}_{13}|\Lambda_{2}|^{2} + (c^{2}_{13}-1)|\Lambda_{3}|^{2} +
  c^{2}_{13}P^{2\nu}_{e1}(|\Lambda_{2}|^{2}-|\Lambda_{1}|^{2})\, ,
\end{equation}
where the oscillation probabilities from $\nu_e$ to the neutrino mass eigenstates $\nu_i$ have been approximated to~\cite{Grimus:2002vb}
\begin{equation}
P^{3\nu}_{e3}  = \sin^2\theta_{13}, \quad 
P^{3\nu}_{e1}  = \cos^2\theta_{13}P^{2\nu}_{e1}, \quad 
P^{3\nu}_{e2}  = \cos^2\theta_{13}P^{2\nu}_{e2}, \quad 
\end{equation}
and the unitarity condition, $P^{2\nu}_{e1}+P^{2\nu}_{e2} = 1$, has been assumed~\footnote{Note our Eq. (\ref{eq:nmm_sun}) differs from Eq. (7) of Ref.~\cite{Borexino:2017fbd}.}.  
Notice that, in this case, Eq.~(\ref{eq:nmm_sun}) has no dependence on any phase, since solar electron neutrinos undergo flavor oscillations arriving to the detector as an incoherent admixture of mass eigenstates. 
In the recent analysis reported by the Borexino collaboration~\cite{Borexino:2017fbd}, the following 90\% C.L. bound on the effective neutrino magnetic moment was reported: $\mu^{M}_{\nu,\,\text{sol}} <2.8\times10^{-11}\mu_{B}$. This constraint  can be directly translated into a limit on the TMM parameters $\left \vert \Lambda_i \right \vert$, as presented in Fig.~\ref{fig:Li_vs_Lj_Borexino}. There, we show  the corresponding 90\% C.L. bounds in the two-dimensional  ($\left \vert \Lambda_i \right \vert$,$\left \vert \Lambda_j \right \vert$) plane when the third element $\left \vert \Lambda_k \right \vert$ is set to zero. 

%
%

Before closing, it is worth mentioning that the effective neutrino magnetic moments can be also studied in other rare-event experiments. 
This is well-motivated by the improved precision expected in the next generation of oscillation and dark matter direct detection experiments, see Ref.~\cite{Huang:2018nxj}. 
In this framework, interesting neutrino sources such as geoneutrinos, atmospheric neutrinos and diffuse supernova background neutrinos that 
contribute to the ``neutrino floor'' at dark matter detectors can be envisaged. They would be expected to provide complementary information 
on neutrino electromagnetic properties. 

\section{Summary and Conclusions}
\label{sec:conclusions}

\begin{table}[t]
\begin{tabular}{@{}lccc@{}}
\toprule
\textbf{Experiment}       & $\left \vert \Lambda_1 \right \vert$ & $\left \vert \Lambda_2 \right \vert$ & $\left \vert \Lambda_3 \right \vert$  \\ \midrule 
\multicolumn{4}{c}{\textbf{SNS prompt}}\\
CsI[Na]   & 69.2 [5.0]      & 70.2  [5.1]       & 89.6  [6.4]    \\
HPGe      & 25.9 [5.1]      & 26.2  [5.2]       & 33.5  [6.6]    \\
LAr       & 14.7 [2.9]      & 14.9  [2.9]       & 19.1  [3.7]    \\
NaI[Tl]   & 16.6 [2.8]      & 16.8  [2.8]       & 21.5  [3.6]    \\ 
\multicolumn{4}{c}{\textbf{SNS delayed}}\\
CsI[Na]   &  54.5 [4.2]      & 48.7  [3.7]       & 49.8 [3.7]    \\
HPGe      &  21.3 [4.2]      & 18.9  [3.8]       & 19.1 [3.8]    \\
LAr       &  11.3 [2.3]      & 10.1  [2.1]       & 10.4 [2.1]    \\
NaI[Tl]   &  10.0 [2.3]      & ~9.1   [2.0]      & ~9.4  [2.0]    \\ 
\multicolumn{4}{c}{\textbf{Reactor}}\\
CONUS     & ~1.9 [0.37]     & ~1.3 [0.26]       & ~1.1  [0.22] \\
CONNIE    & 0.90 [0.13]     & 0.63 [0.09]       & 0.53  [0.08] \\
MINER     & ~1.7 [0.58]     & ~1.2 [0.41]       & ~1.0  [0.34] \\
TEXONO    & ~3.2 [0.46]     & ~2.3 [0.32]       & ~1.9  [0.27] \\
RED100    & ~1.0 [0.14]     & 0.72 [0.10]       & 0.61  [0.08] \\
\multicolumn{4}{c}{\textbf{Solar}}\\
Borexino  & 0.44            & 0.36              & 0.28  \\
\bottomrule 
\end{tabular}
\caption{90\% C.L. limits on TMM elements $\left \vert \Lambda_i \right \vert$, in units of $10^{-10}$ $\mu_B$, from  current and future
  CE$\nu$NS experiments.  The numbers in square brackets indicate the attainable sensitivities in the future setups. Results from the solar neutrino experiment Borexino are also included for comparison.}
\label{tab:TMM-constraints}
\end{table}

In this work, we have examined the potential of the current and next generation of coherent elastic neutrino-nucleus scattering experiments in probing neutrino magnetic moment interactions. 
We have performed a detailed statistical analysis to determine the sensitivities on the three elements of the Majorana neutrino transition magnetic moment matrix, 
 $\left \vert \Lambda_i \right \vert$, that follow from low-energy neutrino-nucleus experiments. 
We have used for the first time the \cevns measurement by the COHERENT experiment at the Spallation Neutron Source in order to constrain the Majorana neutrino 
transition magnetic moments.
By assuming the future setup upgrades in Table~\ref{table:exper} we have also presented the expected sensitivities for the next phases of COHERENT using HPGe, LAr and NaI[Tl] detectors, as well as for reactor neutrino experiments such as CONUS, CONNIE, MINER, TEXONO and RED100.
Our results for the current and future  sensitivities on the TMM elements $\left \vert \Lambda_i \right \vert$ are illustrated in Figs.~\ref{fig:Li_1param} to \ref{fig:Li_vs_Lj_reactors} and summarized in Table~\ref{tab:TMM-constraints}.
From the table, one sees that improvements of at least one order of magnitude compared to the current setup might be expected from future \cevns measurements. Indeed, 
our results show that the next generation \cevns experiments has promising prospects to probe TMMs  at the $10^{-11}\, \mu_B$ level at least. 
It follows that upcoming reactor-based \cevns experiments with low-threshold capabilities have the potential to compete with the current limits 
from $\bar{\nu}_e-e$ scattering data derived in Ref.~\cite{Canas:2015yoa} or with the best current limit reported from Borexino, and translated to our 
general parameterization in Sect.~\ref{sec:other-experiments} (see also the last row of Table~\ref{tab:TMM-constraints}).
As a final remark, we comment that, although the results reported in Table~\ref{tab:TMM-constraints} have been obtained under the assumption of real TMMs, we have also discussed the role of the CP violating phases.


\begin{acknowledgments}

This work is supported by the Spanish grants SEV-2014-0398 and FPA2017-85216-P (AEI/FEDER, UE), PROMETEO/2018/165 (Generalitat
Valenciana) and the Spanish Red Consolider MultiDark FPA2017-90566-REDC. OGM has been supported by CONACYT-Mexico under
grant A1-S-23238 and by SNI (Sistema Nacional de Investigadores). MT acknowledges financial support from MINECO through the Ram\'{o}n y Cajal contract RYC-2013-12438.

\end{acknowledgments}

%
\providecommand{\href}[2]{#2}\begingroup\raggedright\endgroup

\end{document}